\documentclass[aps, prd, preprintnumbers, showpacs, twocolumn, nofootinbib, floatfix]{revtex4}
\usepackage{graphicx}

\newcommand{\PRE}[1]{}       

\def\gappeq{\mathrel{\rlap {\raise.5ex\hbox{$>$}}
            {\lower.5ex\hbox{$\sim$}}}}
\def\lappeq{\mathrel{\rlap{\raise.5ex\hbox{$<$}}
            {\lower.5ex\hbox{$\sim$}}}}

\newcommand{\ba}{\begin{array}}
\newcommand{\ea}{\end{array}}
\newcommand{\bd}{\begin{displaymath}}
\newcommand{\ed}{\end{displaymath}}
\newcommand{\be}{\begin{equation}}
\newcommand{\ee}{\end{equation}}
\def\bt{\begin{table}}
\def\et{\end{table}}
\def\bc{\begin{center}}
\def\ec{\end{center}}
\def\bi{\begin{itemize}}
\def\ei{\end{itemize}}
\def\bw{\begin{widetext}}
\def\ew{\end{widetext}}

\def\bea{\begin{eqnarray}}
\def\eea{\end{eqnarray}}
\def\beas{\begin{eqnarray*}}
\def\eeas{\end{eqnarray*}}

\def\gev{~\rm GeV}
\def\N0{\widetilde{\chi}^0}
\def\Dp{\widetilde{\Delta}^{++}}
\def\Dm{\widetilde{\Delta}^{--}}
\def\Dt{\widetilde{\Delta}}
\def\Cp{\widetilde{\chi}^+}
\def\Cm{\widetilde{\chi}^-}
\def\Cpm{\widetilde{\chi}^\pm}

\def\sel{\widetilde{e}_{L}}
\def\ser{\widetilde{e}_{R}}
\def\sml{\widetilde{\mu}_{L}}
\def\smr{\widetilde{\mu}_{R}}
\def\stl{\widetilde{\tau}_{1}}
\def\str{\widetilde{\tau}_{2}}

\def\D{\Delta}

\def\slash {\!\!\!\!/}

\begin{document}

\preprint{DESY-08-060, IZTECH-P-08-03, CUMQ-HEP-149, HIP-2008-15/TH}
\title{Signals of Doubly-Charged Higgsinos at the CERN Large Hadron Collider }

\author{Durmu\c{s} A. Demir$^{1,2}$} 
\author{Mariana Frank$^3$}
\author{Katri Huitu$^4$} 
\author{Santosh Kumar Rai$^4$}
\author{Ismail Turan$^3$}
\affiliation{$^1$Department of Physics, Izmir Institute of Technology, IZTECH, 
TR35430 Izmir, Turkey.}
\affiliation{$^2$Deutsches Elektronen - Synchrotron, DESY, D-22603 Hamburg, 
Germany.}
\affiliation{$^3$Department of Physics, Concordia University, 
7141 Sherbrooke Street West, Montreal, Quebec, CANADA H4B 1R6.} 
\affiliation{$^4$Department of Physics, University of Helsinki and 
Helsinki Institute of Physics, P.O. Box 64, FIN-00014 University of Helsinki, 
Finland.}
\begin{abstract}
Several supersymmetric models with extended gauge structures, motivated by either grand
unification or by neutrino mass generation, predict light doubly-charged
Higgsinos. In this work we study productions and decays of doubly-charged 
Higgsinos present in left-right supersymmetric models, and show that 
they invariably lead to novel collider signals not found in the minimal
supersymmetric model (MSSM) or in any of its extensions motivated by the 
$\mu$ problem or even in extra dimensional theories. We investigate their 
distinctive signatures at the Large Hadron Collider (LHC) in both pair-- and 
single--production modes, and show that they are powerful tools in determining 
the underlying model via the measurements at the LHC experiments. 
\end{abstract}
\maketitle
\section{Introduction}\label{sec:intro}

The LHC, the highest energy collider ever built starts operating this year, 
and will provide a clean window into `new physics' at the TeV scale. The `new
physics' scenarios, designed to solve the gauge hierarchy problem, generically
bring about new particles and interaction schemes. Supersymmetric theories (SUSY), 
for instance, provide an elegant solution to the gauge hierarchy problem 
by doubling the particle spectrum of the standard model, and their gauge 
sector could be minimal as in the MSSM or non-minimal as in models with 
extended gauge structures. Experiments at the LHC will be probing these 
new  particles as well as new interaction laws among them. 

A general, although not universally present, feature of SUSY, is that, if R-parity $R=(-1)^{(3B+L+2S)}$ (with 
$B$, $L$ and $S$ being baryon, lepton and spin quantum numbers, respectively) is conserved, 
the absolute stability of the lightest supersymmetric partner (LSP) is  guaranteed. This state qualifies to be 
a viable candidate for cold dark matter in the universe (see, for instance, \cite{Demir:2006ef} and
references therein). Supersymmetric models provide a viable dark matter candidate in the lightest
neutral fermion composed of neutral gauginos and Higgsinos. In general, decays of all supersymmetric partners 
necessarily end with the LSP, and given its absolute stability, it leaves any particle detector 
undetected, and thus, appears as a `momentum imbalance' or `missing energy' in collider processes, 
including the ones at the LHC \cite{susysearch}. In this sense, all scattering processes involving
the superpartners are inherently endowed with incomplete final states. 

Though supersymmetry, as an organizing principle, resolves the gauge hierarchy problem, there
is no unique supersymmetric field theory to model `new physics' at the TeV scale. Indeed,
MSSM, though it stands as the minimal supersymmetric model directly constructed from the SM
spectrum, suffers from the well-known $\mu$ problem and lacks a natural understanding of
neutrino masses in the absence of right-handed neutrinos (which must be either ultra-heavy
to facilitate see-saw mechanism or must possess naturally suppressed Yukawa couplings to 
left-handed ones). These features generically require a non-trivial extension of the MSSM
which typically involves additive, or embedding of, gauge structures beyond that of the MSSM. Indeed,
low-energy models following from SUSY GUTs or strings \cite{gut-string} generically predict 
either extension of the SM gauge group by some extra gauge factors, such as a number of 
extra $U(1)$ symmetries, or embedment of the SM gauge group into larger gauge groups. Concerning
the latter, one can consider several structures, for instance, the left-right symmetric SUSY (LRSUSY)
gauge theory $SU(3)_C \times SU(2)_L \times SU(2)_R \times U(1)_{B-L}$. In general, models of `new physics'
(in terms of their gauge and Higgs sectors) are distinguished by certain characteristic signatures
in regard to their lepton and jet spectra of the final state. In this work, we investigate signatures 
specific to LRSUSY and compare with those of the MSSM wherever appropriate. LRSUSY presents
an attractive alternative/generalization of the MSSM  \cite{history, Francis:1990pi}. It 
can be viewed as an alternative to the MSSM by itself or as a covering structure of the MSSM  following
from SUSY GUTs or strings, such as $SO(10)$. LRSUSY models  disallow explicit  R-parity breaking in the Lagrangian, 
thus predicting naturally a supersymmetric dark matter candidate \cite{Demir:2006ef}. They provide a solution to 
the strong and weak CP problems present in the MSSM \cite{Mohapatra:1995xd}. If one chooses Higgs triplet fields, 
with quantum numbers $B-L=\pm 2$, to break the  $SU(2)_R$ gauge group, the neutrino masses turn out to be
induced by the see-saw mechanism \cite{Mohapatra:1979ia}.  The fermionic partners of the Higgs triplet bosons 
are specific to the supersymmetric version, and some of them carry double charge and two units of $L$ number, 
making them perfect testing/searching grounds for exotics. It has been shown that, if the scale for 
left-right symmetry breaking is chosen so that the light neutrinos have the experimentally expected masses, 
these doubly-charged Higgsinos can be light, with masses in the range of ${\cal O} (100)$ GeV
\cite{Aulakh:1997fq,Aulakh:1998nn,Chacko:1997cm}. Such particles could be produced in abundance at the LHC and thus give definite 
identifiable signatures of left-right symmetry. The doubly-charged Higgsinos have been studied in some detail in 
the references \cite{Huitu:1995bc,Raidal:1998vi,Frank:2007nv}, where the production of Higgsinos at 
linear colliders  was analyzed. The doubly-charged Higgsinos can also appear in the so-called 
3-3-1 models (models based on the $SU(3)_c \times SU(3)_L \times U(1)_{N}$ symmetry) \cite{Singer:1980sw}.

In this work, we study doubly-charged Higgsinos at the CERN Large Hadron Collider (LHC), produced singly
or in pair, via various leptonic final states.  We focus on three benchmark points of the model
and analyze the LHC signals resulting from the decays of the doubly-charged Higgsinos. In order to obtain 
definitive predictions for the signal, we perform the analysis in the context of the LRSUSY model, 
though we expect the results for the 3-3-1 model to be similar. The paper is organized as follows: 
In Sec. \ref{sec:lrsusy} we present a brief introduction to the model, for completeness and clarification 
of the notation. In Sec.  \ref{sec:results} we focus on the details and characteristics of the production 
cross sections of the doubly--charged Higgsinos, and discuss their possible decay channels (either through two-body or three-body, depending on the 
spectrum characteristics) proceeding with charged states. Herein we analyze single-- and pair--production 
modes separately.  Finally, in Sec. \ref{sec:conclusion} we conclude and discuss the significance of the 
results in regard to measurements at the LHC.

\section {The Left--Right Supersymmetric Model}
\label{sec:lrsusy}
 
In this section, we review briefly the relevant features of the model necessary for the analysis which follows in the later sections. 
For a more detailed information about the model see, for instance, \cite{history, Francis:1990pi}. The chiral matter
in LRSUSY consist of three families of quark and lepton superfields:
\begin{eqnarray}
\!\!Q\!\!&=&\!\!\left (\begin{array}{c}
u\\ d \end{array} \right ) \sim \left (3,2, 1, \frac13 \right ),
Q^c\!=\!\left (\begin{array}{c}
d^c\\u^c \end{array} \right ) \sim \left ( 3^{\ast},1, 2, -\frac13
\right ),\nonumber \\
L&=&\left (\begin{array}{c}
\nu\\ e\end{array}\right ) \sim\left ( 1,2, 1, -1 \right ),
L^c=\left (\begin{array}{c}
e^c \\ \nu^c \end{array}\right ) \sim \left ( 1,1, 2, 1 \right ),\nonumber
\end{eqnarray}
where the numbers in the brackets denote the quantum numbers under
$SU(3)_C \times SU(2)_L \times SU(2)_R \times U(1)_{B-L}$ gauge factors.

The symmetry breaking is achieved by a Higgs
sector consisting of  bidoublet and triplet Higgs superfields. 
The choice of the triplet Higgs fields has the advantage that it facilitates the 
see--saw mechanism for neutrino masses with renormalizable couplings. Here are the 
decompositions of the Higgs superfields:
\begin{eqnarray}
\label{eq:higgs-decomp}
&&\Phi_1 = \left (\begin{array}{cc}
\Phi^0_{11}&\Phi^+_{11}\\ \Phi_{12}^-& \Phi_{12}^0
\end{array}\right) \sim \left (1,2,2,0 \right),\nonumber \\ 
&&\Phi_2=\left (\begin{array}{cc}
\Phi^0_{21}&\Phi^+_{21}\\ \Phi_{22}^-& \Phi_{22}^0
\end{array}\right) \sim \left (1,2,2,0 \right), \nonumber \\
&&\Delta_{L}=\left(\begin{array}{cc}
\frac {1}{\sqrt{2}}\Delta_L^-&\Delta_L^0\\
\Delta_{L}^{--}&-\frac{1}{\sqrt{2}}\Delta_L^-
\end{array}\right) \sim (1,3,1,-2),\nonumber \\
&&\delta_{L}  =
\left(\begin{array}{cc}
\frac {1}{\sqrt{2}}\delta_L^+&\delta_L^{++}\\
\delta_{L}^{0}&-\frac{1}{\sqrt{2}}\delta_L^+
\end{array}\right) \sim (1,3,1,2),\nonumber \\
&&\Delta_{R} =
\left(\begin{array}{cc}
\frac {1}{\sqrt{2}}\Delta_R^-&\Delta_R^0\\
\Delta_{R}^{--}&-\frac{1}{\sqrt{2}}\Delta_R^-
\end{array}\right) \sim (1,1,3,-2),\nonumber \\
&&\delta_{R}  =
\left(\begin{array}{cc}
\frac {1}{\sqrt{2}}\delta_R^+&\delta_R^{++}\\
\delta_{R}^{0}&-\frac{1}{\sqrt{2}}\delta_R^+
\end{array}\right) \sim (1,1,3,2),
\end{eqnarray}
where numbers in the brackets again denote the quantum numbers of 
fields under $SU(3)_C \times SU(2)_L \times SU(2)_R \times U(1)_{B-L}$.

The superpotential of the model is given by  
\begin{eqnarray}
\label{superpotential}
W & = & {\bf Y}_{Q}^{(i)} Q^T\Phi_{i}i \tau_{2}Q^{c} + {\bf Y}_{L}^{(i)}
L^T \Phi_{i}i \tau_{2}L^{c} \nonumber\\
&&+ i({\bf h}_{ll}L^T\tau_{2} \delta_L L +
{\bf h}_{ll}L^{cT}\tau_{2}
\Delta_R L^{c}) \nonumber \\
&& + \mu_{3} \mbox{Tr}\left[ \Delta_L  \delta_L +\Delta_R 
\delta_R\right] 
+ \mu_{ij}\mbox{Tr}\left[i\tau_{2}\Phi^{T}_{i} i\tau_{2} \Phi_{j}\right]\nonumber\\
&&+W_{NR}, 
\end{eqnarray}
where $W_{NR}$ denotes (possible) non-renormalizable terms arising from integrating-out
of the heavier fields. The Lagrangian of the model, as usual, consists of 
the standard $F$-terms, $D$-terms as well as the soft SUSY--breaking terms. Considering
the decay and production processes under investigation, the relevant parts of the 
soft--breaking Lagrangian read as 
\begin{eqnarray}
\label{eq:soft}
-L_{soft}&=&({m}_{\Phi}^{2})_{ij} \Phi_i^{\dagger}  \Phi_j + (m_{L}^2)_{ij}{\tilde l}_{Li}^{\dagger}{\tilde
l}_{Lj}+ (m_{R}^2)_{ij}{\tilde l}_{Ri}^{\dagger}{\tilde l}_{Rj}\nonumber\\
&+& \left[{\bf A}_{L}^{i}{\bf Y}_{L}^{(i)}{\tilde L}^T \Phi_{i}
i\tau_{2}{\tilde L}^{c}\right. \nonumber \\
&& \left. + i{\bf A}_{LR} {\bf h}_{ll}\left({\tilde L}^T\tau_{2}
\delta_L{\tilde  L} + {\tilde L}^{cT}\tau_{2} \Delta_R{\tilde L}^{c}\right)
+ h.c.\right] 
\nonumber\\ 
&-& \left[M_{LR}^2 \mbox{Tr}\left[
\Delta_R  \delta_R + \Delta_L 
 \delta_L \right]\right.\nonumber\\
&-& \left.[B \mu_{ij} \Phi_{i} \Phi_{j}+h.c.\right]
\end{eqnarray}
where the first line stands for mass-squared terms, the second and third for trilinear couplings
(holomorphically corresponding to similar terms in (\ref{superpotential})),
 and the last two for bilinear couplings.

Combining (\ref{eq:soft}) with $F$-term and $D$-term contributions, minimization of the 
Higgs potential gives vacuum expectation values (VEVs) for the neutral components of the Higgs 
fields in (\ref{eq:higgs-decomp}), as discussed in detail in  \cite{Demir:2006ef,Frank:2007nv}.

In the following, we give a detailed discussion of the charged and neutral fermions as well as 
sleptons in LRSUSY in preparation for a thorough analysis of the productions and decays of the 
doubly-charged Higgsinos. 

\subsection{Charginos}

As follows from the decompositions of the Higgs fields in (\ref{eq:higgs-decomp}),
the particle spectrum consists of doubly-charged Higgsinos $\tilde \Delta_L^{--}, \tilde \delta_L^{++},\tilde
\Delta_R^{--}$ and $\tilde \delta_R^{++}$. In the Lagrangian basis they possess the bilinear terms
\begin{equation}
{\cal L}_{\tilde \Delta}=-M_{\tilde \Delta^{--}}\tilde \Delta_L^{--}\tilde \delta_L^{++}-M_{\tilde \Delta_R^{--}}\tilde 
\Delta_R^{--}\tilde\delta_R^{++} + {h.c.}\,, 
\end{equation}
where the Higgsino mass $M_{\tilde \Delta^{--}} \equiv \mu_3$ in the notation of (\ref{superpotential}).
In addition to these doubly-charged ones, the model consists also a total of six singly-charged Higgsinos
and gauginos, corresponding to $\lambda_{L}$, $\lambda_{R}$, $\tilde\phi_{u}$,
$\tilde\phi_{d}$, $\tilde\Delta_{L}^{-}$, $\tilde\delta_{L}^{+}$, $\tilde\delta_{R}^{+}$ and $\tilde\Delta_{R}^{-}$. The bilinears in
these charged states combine to give 
\begin{equation}
{\cal L}_C=-\frac{1}{2}(\psi^{+T}, \psi^{-T}) \left ( \begin{array}{cc}
                                                        0 & X^T \\
                                                        X & 0
                                                      \end{array}
                                              \right ) \left (
\begin{array}{c}
                                                               \psi^+ \\
                                                               \psi^-
                                                               \end{array}
                                                        \right ) + {\rm h.c.} \ , 
\end{equation}
where $\psi^{+T}=(-i \lambda^+_L, -i \lambda^+_R, \tilde{\phi}_{1d}^+,
\tilde{\phi}_{1u}^+, \tilde{\delta}_L^+, \tilde{\delta}_R^+)$,
$\psi^{-T}=(-i \lambda^-_L, -i \lambda^-_R, \tilde{\phi}_{2d}^-,
\tilde{\phi}_{2u}^-, \tilde{\Delta}_L^-, \tilde{\Delta}_R^-)$, and
\begin{eqnarray}
\small 
X\!\!\!=\!\!\!\left( \begin{array}{cccccc}
                            M_L & 0 & 0 & g_L\kappa_d & \sqrt{2}g_Lv_{\delta_L} &0
\\
           0 & M_R &0  & g_R\kappa_d &0
&\sqrt{2}g_Rv_{\delta_R}
\\
    g_L\kappa_u & g_R \kappa_u & 0 &-\mu_1 &  0 &0
\\
0 & 0  & -\mu_1 &0 & 0&0\\
 \sqrt{2} g_L v_{\Delta_L} & 0 & 0 &0 & -\mu_3 &0 \\
       0 & \sqrt{2} g_R v_{\Delta_R} & 0 & 0 &0& -\mu_3
               \end{array}
         \right)\nonumber
\end{eqnarray}
in the mass mixing matrix. We have set, for simplicity, 
$\mu_{ij}\equiv \mu_1$ for all $(i\neq j)$. Here 
$\kappa_u = \langle \Phi^0_{11} \rangle$, $\kappa_d = 
\langle \Phi^0_{22} \rangle$, $v_{\Delta_{L,R}} = \langle \Delta_{L,R}^0 \rangle$,
$v_{\delta_{L,R}} = \langle \delta^0_{L,R} \rangle$, and 
$M_{L,R}$ are the $SU(2)_{L,R}$ gaugino masses,
respectively. The physical chargino states $\tilde{\chi}_i$ are obtained by
\begin{eqnarray}
\tilde{ \chi}_i^+=V_{ij}\psi_j^+, \ \tilde{ \chi}_i^-=U_{ij}\psi_j^- \;\; \left(i,j=1, \ldots 6\right),
\end{eqnarray}
with $V$ and $U$ unitary matrices satisfying
\begin{equation}
U^* X V^{-1} = M_D
\end{equation}
where $M_D$ is a $6\times 6$ diagonal matrix with non-negative entries.
The mixing matrices $U$ and $V$ are obtained by computing the eigensystem
of  $X X^{\dagger}$ and $X^{\dagger} X$, respectively.

While $\kappa_u$ and $\kappa_d$ are the VEVs responsible for giving masses to quarks and leptons, the non-MSSM Higgs VEVs,
$v_{\delta_L}$ and $ v_{\Delta_R}$ are responsible for neutrino
masses.   $v_{\Delta_L}$ and $v_{\delta_L}$
enter in the formula for the mass of $W_L$ (or the $\rho$
parameter), while $v_{\Delta_R}, v_{\delta_R}$ enter in the
formula for the mass of $W_R$. It is thus justified to take
$v_{\Delta_L}, v_{\delta_L}$ to be negligibly small.  For
$v_{\Delta_R} $ there are two possibilities: either  $v_{\Delta_R}
$ is $\approx 10^{13}$ GeV \cite{Frank:2002hk, Aulakh:1997fq},
which supports the seesaw mechanism, leptogenesis and provides
masses for the light neutrinos in agreement with experimental
constraints, but offers no hope to see right-handed particles; or
$v_{\Delta_R} $ is $\approx 1- 10$ TeV, but one must introduce
something else (generally an intermediate scale, or an extra
symmetry) to make the neutrinos light  \cite{Aulakh:1997fq,Aulakh:1998nn,Aulakh:1999pz}.

\subsection{Neutralinos}
\label{subsec:neutralinos}
In LRSUSY there are  eleven neutral fermions, corresponding to
$\lambda_{Z}$,
$\lambda_{Z^{\prime}}$,
$\lambda_{B-L}$,  $\tilde\phi_{1u}^0$, $\tilde\phi_{2u}^0$,
$\tilde\phi_{1d}^0$,  $\tilde\phi_{2d}^0$, $\tilde\Delta_{L}^0$,
$\tilde\Delta_{R}^0$,  $\tilde\delta_{L}^0$ and $\tilde\delta_{R}^0$. 
Their bilinears give the contribution to the Lagrangian 
\begin{equation}
{\cal L}_N=-\frac{1}{2} {\psi^0}^T Z \psi^0  + {\rm h.c.} \ ,
\end{equation}
where $\psi^0=(-i \lambda_L^0, -i \lambda_R^0, -i \lambda_{B-L},
\tilde{\phi}_{1u}^0, \tilde{\phi}^0_{2d}, \tilde{\Delta}_L^0,
\tilde{\delta}_L^0, \break\tilde{\Delta}_R^0, \tilde{\delta}_R^0, \tilde{\phi}_{1d}^0, \tilde{\phi}^0_{2u} )^T$,
and the mass mixing matrix $Z$ is given by 
\begin{widetext}
\begin{equation}
\displaystyle
Z=\left(\!\! \begin{array}{ccccccccccc}
              \!M_{L} &\! 0 & \!0 &\! -\frac{g_L \kappa_u}{\sqrt{2}} &\!
\frac{g_L \kappa_d}{\sqrt{2}} &\! -2^{\frac12}g_Lv_{\Delta_L}&\! -2^{\frac12}g_Lv_{\delta_L}&\!0 &\! 0&\! 0&\! 0 
\\
         \!0 & \!M_{R} &\! 0 &\! \frac{g_L \kappa_u}{\sqrt{2}} &\! \frac{g_L
\kappa_d}{\sqrt{2}} &\! 0 &\!0 &\! -2^{\frac12}g_Rv_{\Delta_R} &\! -2^{\frac12}g_R v_{\delta_R}&\! 0&\! 0
 \\
           \!0 \!& \!0 & \!M_{B-L} & \!0 & \!0 & \!2^{\frac32}g_V v_{\Delta_L} & \!2^{\frac32} g_V
v_{\delta_L} &\!2^{\frac32}g_V v_{\Delta_R} &\! 2^{\frac32} g_V
v_{\delta_R}&\! 0&\! 0
\\
        \! -\frac{g_L \kappa_u}{\sqrt{2}}&\! \frac{g_R \kappa_u}{\sqrt{2}}
&\! 0 &\! 0 & \!\mu_1 &\! 0 &\! 0 &\! 0 &\! 0&\! 0&\! 0  
\\
             \! \frac{g_L \kappa_d}{\sqrt{2}} &\! -\frac{g_R \kappa_d}{\sqrt{2}} & \!0 &\!\mu_1
& \!0 & \!0 &\! 0 & \!0 & \!0&\! 0&\! 0
  \\
\!-2^{\frac12}g_Lv_{\Delta_L} & \!0 &\!2^{\frac32}g_Vv_{\Delta_L}&\! 0&\! 0&\!0 &\!-\mu_3 &\! 0 &\!0&\! 0&\! 0 \\
  \!-2^{\frac12}g_L v_{\delta_L} &\! 0& \! 2^{\frac32}g_V v_{\delta_L} & \!0  &\!0 &\! -\mu_3 & \!0 &\!0 &\!0&\! 0&\! 0
  \\
             \!0&\!-2^{\frac12}g_Rv_{\Delta_R} &\!2^{\frac32}g_Vv_{\Delta_R}& \!0&\! 0&\! 0 & \!0 &\!0 &\!-\mu_3 &\! 0&\! 0
\\
\! 0 &\! -2^{\frac12}g_R v_{\delta_R} &\! 2^{\frac32}g_V v_{\delta_R} & \!0 & \!0&\! 0 &\! 0 &\! -\mu_3 &\! 0&\! 0&\! 0
\\
\! 0&\! 0&\! 0&\! 0&\! 0&\! 0&\! 0&\! 0&\! 0&\! 0&\!\mu_1
\\
\! 0&\! 0&\! 0&\! 0&\! 0&\! 0&\! 0&\! 0&\! 0&\!\mu_1&\! 0 
       \end{array}\!\! \!
        \right )
\end{equation}
\end{widetext}
with $M_{B-L}$ being the $U(1)_{B-L}$ gaugino mass. The physical neutralinos are defined via
\begin{equation}
\tilde{\chi}^0_i=N_{ij} \psi^0_j \ ~~ (i,j=1,2, \ldots 11),
\end{equation}
where $N$ is the unitary matrix that diagonalizes $Z$
\begin{equation}
N^* Z N^T = Z_D,
\label{equationN}
\end{equation}
with $Z_D$ being a $11\times 11$ diagonal matrix with non-negative entries. The lightest of 
the eleven neutralinos, $\tilde\chi_1^0$, is a candidate for cold dark matter in the universe
\cite{Demir:2006ef}.

\subsection{Scalar leptons}
\label{subsec:sleptons}
Combining $F$-term, $D$--term and soft-breaking contributions pertaining
to sleptons, their mass-squared matrix is found to be 
\begin{eqnarray}
\displaystyle
{\cal M}_{L}^2 = \left( \begin{array}{c@{\hspace*{0.4cm}}c}
M^2_{LL} & M^2_{LR} \\\\
M^2_{RL} & M^2_{RR}
                         \end{array}
                  \right)
\end{eqnarray}
where 
\begin{eqnarray}
M^2_{LL} &=& M_{L}^2+ m_{\ell}^2+m_Z^2(T_{3\ell}+ \sin^2 \theta_{\rm W}) \cos 2\beta,
\nonumber\\
M^2_{LR} &=& M_{RL}^{2\, \dagger} = m_{\ell} (A+\mu \tan \beta),
\nonumber\\
M^2_{RR} &=& M_{R}^2+ m_{\ell}^2-m_Z^2  \sin^2 \theta_{\rm W} \cos2 \beta
\end{eqnarray}
as follows from (\ref{eq:soft}) with $\ell =e, \mu, \tau$. We neglect
intergenerational mixings, and intragenerational left-right mixing 
can be important only for $\ell=\tau$ flavor.

\section{Production and Decay of Doubly-Charged Higgsinos}
\label{sec:results}
\begin{center}
\begin{figure}[t]
	\includegraphics[width=3.4in]{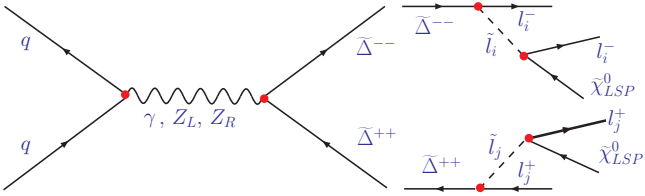} 
\caption{\sl\small  
Direct production of $\widetilde{\Delta}^{--}$ pair at the LHC. 
Subsequent decays of $\widetilde{\Delta}^{--}$ give rise to two dileptons plus 
missing energy signal, if $M_{\;\widetilde{l}_j}<M_{\widetilde{\;\Delta}^{--}}$.}
\label{fig:pairprod}
\end{figure}
\end{center}

\begin{center}
\begin{figure}[t]
	\includegraphics[width=3.45in]{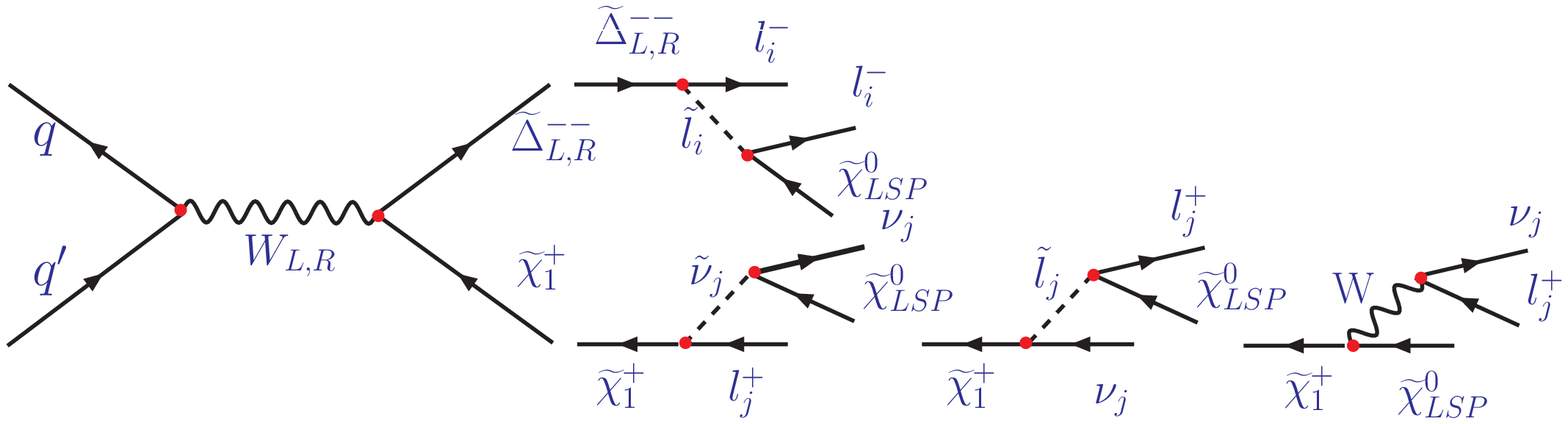} 
\caption{\sl\small 
Direct production of single $\widetilde{\Delta}^{--}$ in association with 
$\widetilde{\chi}_1^+$ at the LHC. Subsequent decays of 
$\widetilde{\Delta}^{--}$ and $\widetilde{\chi}_1^+$ give rise to a trilepton 
plus missing energy signal, if $M_{\;\widetilde{\nu}_j} < M_{\widetilde{\chi}_1^+}$ 
and $M_{\;\widetilde{l}_j}< M_{\widetilde{\;\Delta}^{--}}$.}
\label{fig:singleprod}
\end{figure}
\end{center}

Having described neutralino, chargino and slepton sectors in detail, we now 
analyze productions and decays of doubly-charged Higgsinos. The relevant 
Feynman rules are listed in the Appendix. The pair--production processes 
at the LHC involve
\bi
\item $p\, p \longrightarrow \Dp\, \Dm $ (illustrated in Fig.~\ref{fig:pairprod})
\ei 
which proceeds with $s$-channel $\gamma$ and $Z_{L,R}$ exchanges, and
\bi
\item $p\, p \longrightarrow \Cp_1\, \Dm $ (illustrated in Fig.~\ref{fig:singleprod})
\ei
which rests on $s$-channel $W_{L,R}$ exchanges. Both processes are 
generated by quark--anti-quark annihilation at the parton level. 
The $s$-channel Higgs exchanges cannot give any significant contribution.

These doubly-- and singly--charged fermions  subsequently 
decay via a chain of cascades until the lightest neutralino $\chi_1^0$ 
is reached. Given that charged leptons ($\ell=e$ and $\ell=\mu$, 
especially) give rise to rather clean signals at the ATLAS and CMS
detectors, we classify final states according to their lepton 
content in number, electric charge and flavor. In general, the two-body 
decays of doubly-charged Higgsinos are given by
\bi
\item $\Dm \longrightarrow \widetilde{\ell}^- ~\ell^-$,
\item $\Dm \longrightarrow \Delta^{--} ~\N0_i$,
\item $\Dm \longrightarrow \Cm_i ~\Delta^{-}$,
\item $\Dm \longrightarrow \Cm_i ~W^{-}$,
\ei
whose decay products further cascade into lower-mass 
daughter particles of which leptons are of particular
interest. The production and decay processes 
mentioned here are illustrated in Fig.~\ref{fig:pairprod}
and Fig.~\ref{fig:singleprod}. Clearly, pair-produced
doubly-charged Higgsinos lead to $4 \ell + E\slash_T$
final states whereas single-produced doubly-charged Higgsinos
give rise to $3 \ell + E\slash_T$ signals.

We assume that triplet Higgs bosons are heavier and degenerate in mass, 
which renders them kinematically inaccessible for decay modes of the 
relatively lighter doubly-charged Higgsinos. The possibility of light observable doubly charged Higgs bosons has been explored extensively in both phenomenological analyzes \cite{dcbtheory} and experimental investigations \cite{dcbexperiment} and is beyond the scope of this study. Therefore, we concentrate on the 
remaining accessible decay channels. For the numerical estimates we consider 
three sample points in the LRSUSY parameter space, as tabulated in 
Table \ref{susyin}. A quick look at the resulting mass spectrum for the 
sparticles suggest that the chargino states are also heavier than or comparable to the 
doubly-charged Higgsinos, and hence, the favorable decay channel for 
$\Dt$ is $\Dm \longrightarrow \widetilde{\ell}^- ~ \ell^-$, provided
that $m_{\tilde{l}} < M_{\Dm}$. For relatively light Higgsinos, one can, in
principle, have $m_{\tilde{l}} > M_{\Dm}$ in which case the
only allowed decay mode for the doubly-charged
Higgsinos would be the 3-body decays, which would proceed dominantly through off-shell sleptons:  $\Dm \to
\widetilde{\ell}^{\star\, -} ~ \ell^- \to \ell^- \ell^- \N0_1$. We have explicitly checked that 
the 3-body decay of the doubly-charged Higgsinos through the heavy off-shell 
charginos or $W$ bosons is quite suppressed with respect to the two body decay, 
and can be safely neglected.
\begin{widetext}
\begin{center}
\bt[h]
$$
\begin{array}{c|c|c|c}\hline
           & {\rm \bf SPA}  & {\rm \bf SPB} & {\rm \bf SPC}\\
           &\tan\beta=5,M_{B-L}= 25\gev
           &\tan\beta=5,M_{B-L}= 100\gev
	   &\tan\beta=5,M_{B-L}= 0\gev\\
\mbox{Fields} &M_L=M_R=250\gev
           &M_L=M_R=500\gev
           &M_L=M_R=500\gev \\
           &v_{\D_R}=3000\gev,v_{\delta_R}=1000\gev
           &v_{\D_R}=2500\gev,v_{\delta_R}=1500\gev 
	   &v_{\D_R}=2500\gev,v_{\delta_R}=1500\gev\\ 
	   &\mu_1 = 1000\gev,\mu_3 =300\gev
           &\mu_1 = 500\gev, \mu_3 =500\gev
	   &\mu_1 = 500\gev,\mu_3 =300\gev \\
\hline \hline
\N0_i~(i=1,3) & 89.9, 180.6, 250.9\gev & 212.9,441.2,458.5\gev & 142.5, 265.6, 300.0\gev\\
\Cpm_i~(i=1,3)& 250.9,300.0,953.9\gev & 459.4, 500.0,500.0\gev & 300.0, 459.3, 500.0\gev\\
M_{\Dt}       & 300\gev & 500\gev & 300\gev \\
  W_R, Z_R    & 2090.4, 3508.5\gev    & 1927.2, 3234.8\gev & 1927.2, 3234.8\gev \\ \hline
	      & {\bf S2} ~~~~~~~~~~~~~~~~~~ {\bf S3}
	      & {\bf S2} ~~~~~~~~~~~~~~~~~~ {\bf S3} 
              & {\bf S2} ~~~~~~~~~~~~~~~~~~ {\bf S3}    \\ 
\sel,\ser     & (156.9,155.6\gev), (402,402\gev) & (254.2,253.4\gev), 
(552,552\gev) & (214.9,214.0\gev), (402.6,402.2\gev)\\
\sml,\smr     & (156.9,155.6\gev), (402,402\gev) & (254.2,253.4\gev), 
(552,552\gev) & (214.9,214.0\gev), (402.6,402.2\gev)\\
\stl,\str     & (155.4,159.9\gev), (401,406\gev) & (252.5,257.9\gev), 
(550,556\gev) & (212.8,216.2\gev),(401.5,403.3\gev)\\\hline \hline

\end{array}
$$
\caption{\sl\small The numerical values assigned to the model parameters in 
defining the sample points {\bf SPA}, {\bf SPB} and {\bf SPC}. In each case, 
{\bf S2} and {\bf S3} designate parameter values which allow for {\bf 2}-body 
and {\bf 3}-body decays of doubly-charged Higgsinos, respectively. The VEVs of 
the left-handed Higgs triplets are taken 
as $v_{\D_L}\sim v_{\delta_L}\simeq 10^{-8}\gev$. For the couplings we use
$g_L=g_R=g$ and for $h_{ll}=0.1$ \cite{Frank:2007nv}.}
\label{susyin}
\et
\end{center}
\end{widetext}

We present our results for the Higgsino pair production for the two sample points {\bf SPA} and {\bf SPB} described in Table~\ref{susyin}. Since the 
cross sections for the single production modes are highly suppressed for {\bf SPA} and {\bf SPB}, we consider yet another sample point, called {\bf 
SPC} in Table~\ref{susyin}, which maximizes the single production cross section of $\widetilde{\Delta}_L^{--}$. It is also possible to find a sample point 
which maximizes the cross section for single $\widetilde{\Delta}_R^{--}$ production. We discuss the single $\widetilde{\Delta}_L^{--}$ production in 
detail and comment on the $\widetilde{\Delta}_R^{--}$ case, as their features are fairly similar.

For the benchmark point in Table~\ref{susyin}, the doubly-charged Higgsinos
assume the following  2-- and 3--body decay branchings:
\begin{eqnarray}
\label{eq:decays}
 BR(\Dm_{L/R} \to \tilde{\ell}_{iL/iR}^-\ell_i^-) &\simeq& \frac{1}{3},
~~~m_{\tilde l_{i}}<M_{\tilde\Delta^{--}} \nonumber\\
 BR(\tilde{\ell}_{iL/iR}^- \to \ell_i^- \N0_1) &=& 1, \\
 BR(\Dm_{L/R} \to \ell_i^- \ell_i^-  \N0_1) &\simeq& \frac{1}{3}, 
~~~m_{\tilde l_{i}}>M_{\tilde\Delta^{--}} \nonumber
\end{eqnarray}
where $i=e,\mu,\tau$. One notes that only 3-body decay channel is allowed when 
$m_{\tilde{\ell}_{i}}>M_{\Dt}$. (We discuss the chargino decay later for the single production mode).
To fix our notations, we denote by {\bf S2} the signal corresponding to the 
{\bf 2}-body decay of $\Dt$ and by {\bf S3} the signal corresponding to the
{\bf 3}-body decay of $\Dt$. The two separate cases correspond to two 
different choices of the slepton masses for the same sample point. These features
are shown in parentheses as columns in Table~\ref{susyin} for {\bf SPA}, 
{\bf SPB} and {\bf SPC}.

In what follows we shall analyze single-- and pair--productions of 
doubly-charged Higgsinos separately by using Monte Carlos techniques.

\subsection{Pair-production of doubly-charged Higgsinos}

\begin{figure}[t]
\begin{center}
\hspace*{-0.8 in}\includegraphics[height=2.8in,width=3.4in]{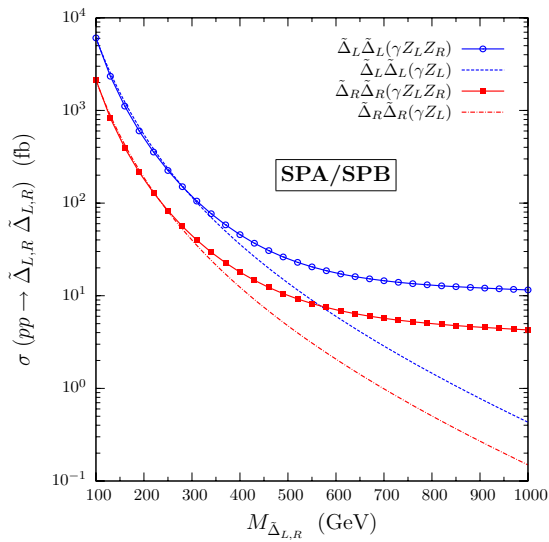} 
\caption{\sl\small The pair-production cross sections for doubly-charged Higgsinos
in LRSUSY at the LHC. The plots are performed by using the parameter sets
{\bf SPA}/{\bf SPB} except that $M_{\tilde \Delta^{--}} \equiv \mu_3$ is 
allowed to vary from $100\ {\rm GeV}$ up to $1\ {\rm TeV}$.
See the text for explanation of curves.}
\label{csec}
\end{center}
\end{figure}

The pair--production of doubly--charged Higgsinos at the LHC occurs through the 
$s$-channel exchanges of the neutral gauge bosons in the model, as depicted in 
Fig.~\ref{fig:pairprod}. The heavy $Z$ boson ($Z_R$) can enhance the production cross section through 
resonance effect, if kinematically accessible at the LHC. In Fig.~\ref{csec} 
we plot production cross sections for $\Dm$ chiralities and exchanged gauge bosons. 
It is seen that cross section is quite sizeable for sufficiently light 
doubly--charged Higgsinos:  it starts at $\sim 10^{4}\ {\rm fb}$ at $M_{\Dt} \simeq
100\ {\rm GeV}$ and stays above $\sim 10\ {\rm fb}$ even if $M_{\Dt}$ is stretched 
up to $1\ {\rm TeV}$ provided that contributions of all three neutral gauge 
bosons, $\gamma$, $Z_L$ and $Z_R$, are included. The figure also shows that cross
sections, for both chirality, fall rapidly with increasing $M_{\Dt}$ if $Z_R$ gauge 
boson is decoupled from the low-energy spectrum. The plots highlight the fact that 
the heavy $Z_R$ contribution becomes more significant for 
pair production of heavier states, as seen in Fig.~\ref{csec}. Pair--production of 
heavier states requires a much higher effective center of mass energy 
$\sqrt{\hat {s}}=\sqrt{x_1 x_2 s}$, where $x_i$'s are the momentum fractions carried by the partons at the 
hadron collider. This would yield a stronger $s$-channel suppression of the SM 
contributions coming from the photon and $Z$ exchange and enhance the 
contribution coming from the heavy $Z_R$ exchange.  

The doubly-charged Higgsinos decay according to Eq.~\ref{eq:decays} into 
two same-sign same-flavor (SSSF) leptons and the lightest neutralino $\N0_1$,
the LSP. This decay pattern gives rise to final states involving four isolated
leptons of the form $\left(\ell_i^- \ell_i^-\right)\, \left(\ell_j^{+} \ell_j^{+}\right)$
where $\ell_i$ and $\ell_j$ are not necessarily identical lepton flavors. More
precisely, final states generated by the decays of doubly-charged Higgsino pairs
generically contain tetraleptons plus missing momentum carried away by the LSP:
\begin{equation}
p p \longrightarrow \Dp \Dm \longrightarrow  \left(\ell^+_i \ell^+_i\right) + 
\left(\ell^-_j \ell^-_j\right) + E\slash_T\,,
\label{eq:pairprod}
\end{equation}
where $\ell_i,\ell_j=e,\mu,\tau$.

The $4\ell+E\slash_T$ signal receives contributions from the pair-production of
both chiral states of the doubly-charged Higgsino. Since at the LHC it is difficult
to determine chiralities of particles, it is necessary to add up 
their individual contributions to obtain the total number 
of events. This yields a rather clean and robust $4l+$ missing $p_T$ 
signal at the LHC with highly suppressed SM background. In fact, one 
finds that the SM background with tetraleptons, where $\ell_i=e$ and $\ell_j=\mu$ in
Eq.~\ref{eq:pairprod} with large missing transverse energy 
($E\slash_T \ge 50\gev$), is very suppressed ($\mathcal{O}\sim 10^{-3}$ fb) 
and can therefore be safely neglected compared to the signal generated by 
doubly-charged Higgsino pairs. This fact makes this channel highly promising 
for an efficient and clean disentanglement of LRSUSY effects. 

For triggering and enhancing the $4\ell+E\slash_T$ signal we impose the 
following kinematic cuts:
\begin{itemize}
\item  The charged leptons in the final state must respect the rapidity cut
$|\eta_{\ell}|<2.5$,
\item  The charged leptons in the final state must have a transverse momentum $p_T > 25\gev$.
\item To ensure proper resolution between the final state leptons we demand
$\D R_{\ell\ell}>0.4$ for each pair of leptons, where $ \D R = \sqrt{ (\D \phi)^2 + (\D \eta)^2 }$,
$\phi$ being the azimuthal angle. 
\item The missing transverse energy must be $E\slash_T > 50$ GeV.
\item The pairs of oppositely-charged leptons of same flavor have at least 
$10\ {\rm GeV}$ invariant mass.
\end{itemize}

For numerical analysis, we have included the LRSUSY model into 
{\tt CalcHEP 2.4.5} \cite{calchep} and generated the event files for the production 
and decays of the doubly-charged Higgsinos using the {\tt CalcHEP} event generator. 
The event files are then passed through the {\tt CalcHEP+Pythia} interface where we include the effects 
of both initial and final state radiations using Pythia switches \cite{pythia} 
to smear the final states. We use the leading order CTEQ6L \cite{cteq}
parton distribution functions (PDF) for the quarks in protons. 

Below we list production cross sections as well as total event cross 
sections ( after applying the kinematic cuts mentioned above). For 
four-lepton plus missing energy signal we take specifically 
$2\mu^- + 2e^+ +E\slash_T$ final state, and find the following results 
for {\bf SPA} and {\bf SPB}:
\begin{itemize}
\item {\bf SPA}: 
$$\sigma(\Dm_{L}\Dp_{L})=117.9~{\rm fb}$$ and 
$$\sigma(\Dm_{R}\Dp_{R})=44.5~{\rm fb}.$$ 
After imposing the kinematic cuts, the total cross section for the final 
state (summing over contributions coming from  doubly-charged Higgsinos of either
chirality) turns out to be:
\bi
\item  {\bf S2} ~ $\sigma(2\mu^-2e^++E\slash_T) = 7.71$ fb,
\item  {\bf S3} ~ $\sigma(2\mu^-2e^++E\slash_T) = 7.02$ fb.
\ei
\item {\bf SPB}: 
$$\sigma(\Dm_{L}\Dp_{L})=32.4~{\rm fb}$$ and
$$\sigma(\Dm_{R}\Dp_{R})=12.95~{\rm fb}.$$
After applying the kinematic cuts we find:
\bi
\item  {\bf S2} ~ $\sigma(2\mu^-2e^++E\slash_T) = 2.43$ fb,
\item  {\bf S3} ~ $\sigma(2\mu^-2e^++E\slash_T) = 2.66$ fb.
\ei
\end{itemize}
The same numerical results hold also when the final state is 
charge-conjugated $i.e.$ $2\mu^+2e^-+E\slash_T$. In principle, one can also 
work with final states where one of the lepton flavors is $\tau$. 
Then one needs to fold in the efficiencies for $\tau$ identification at LHC 
with the above numbers to get the correct event rates. 
\begin{figure}[t]
\begin{center}
\includegraphics[height=2.7in,width=2.7in]{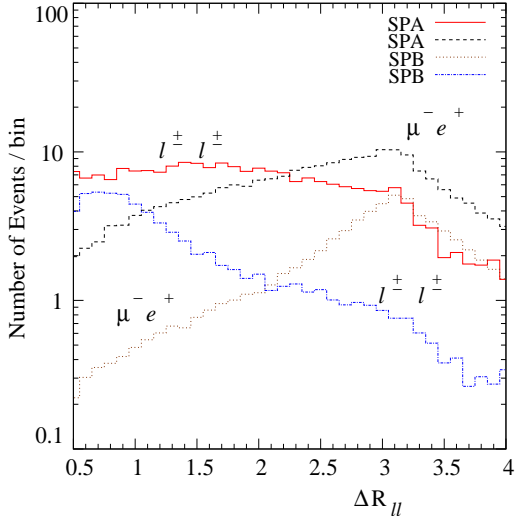} 
\caption{\sl\small Binwise distribution of $\Delta R$ with binsize 0.1
and integrated luminosity of $\int{\cal L} dt= 30 fb^{-1}$.}
\label{DR4l}
\end{center}
\end{figure}

In Fig.~\ref{DR4l} we plot the binwise distribution of the spatial resolution
between the charged lepton pairs for the different cases indicated on the curves. 
We choose to use the events for the case {\bf S2} for both {\bf SPA} 
and {\bf SPB}, as the characteristic features of the distributions remain 
the same for {\bf S3}.
Here the notation $l^\pm l^\pm$ stands for $\mu^-\mu^-$ or $e^+e^+$. The figure manifestly 
shows the difference between the SSSF leptons whose
distributions are peaked at low values of $\D R$ and the opposite-sign
different-flavor (OSDF) leptons whose distributions maximize at higher 
values of $\D R$. The SSSF leptons originate from the cascade decay of 
one single doubly-charged Higgsino whereas OSDF lepton configurations are 
formed by two isolated leptons, one originating from $\Dm$, the other from $\Dp$. 
To this end, SSSF leptons with small spatial separation qualify to 
be a direct indication of the doubly-charged Higgsinos in the 
spectrum (of the LRSUSY or of 3-3-1 model, for example). This feature is 
a clear-cut signal of extended SUSY models as it does not exist in 
the MSSM or in any of its extensions that contain only singly-charged fields.

\begin{figure}[t]
\begin{center}
\includegraphics[height=2.7in,width=2.7in]{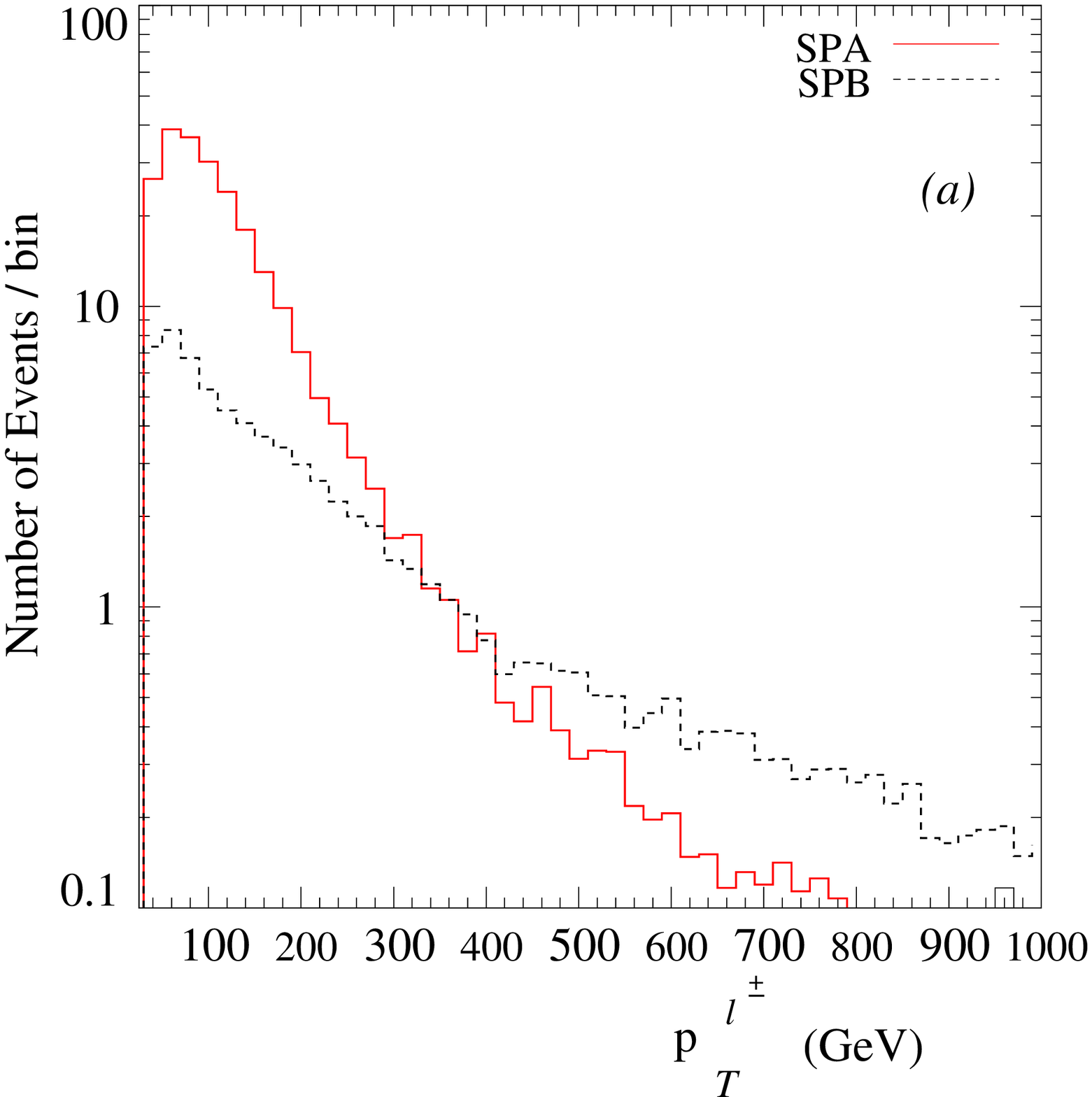} 
\includegraphics[height=2.7in,width=2.7in]{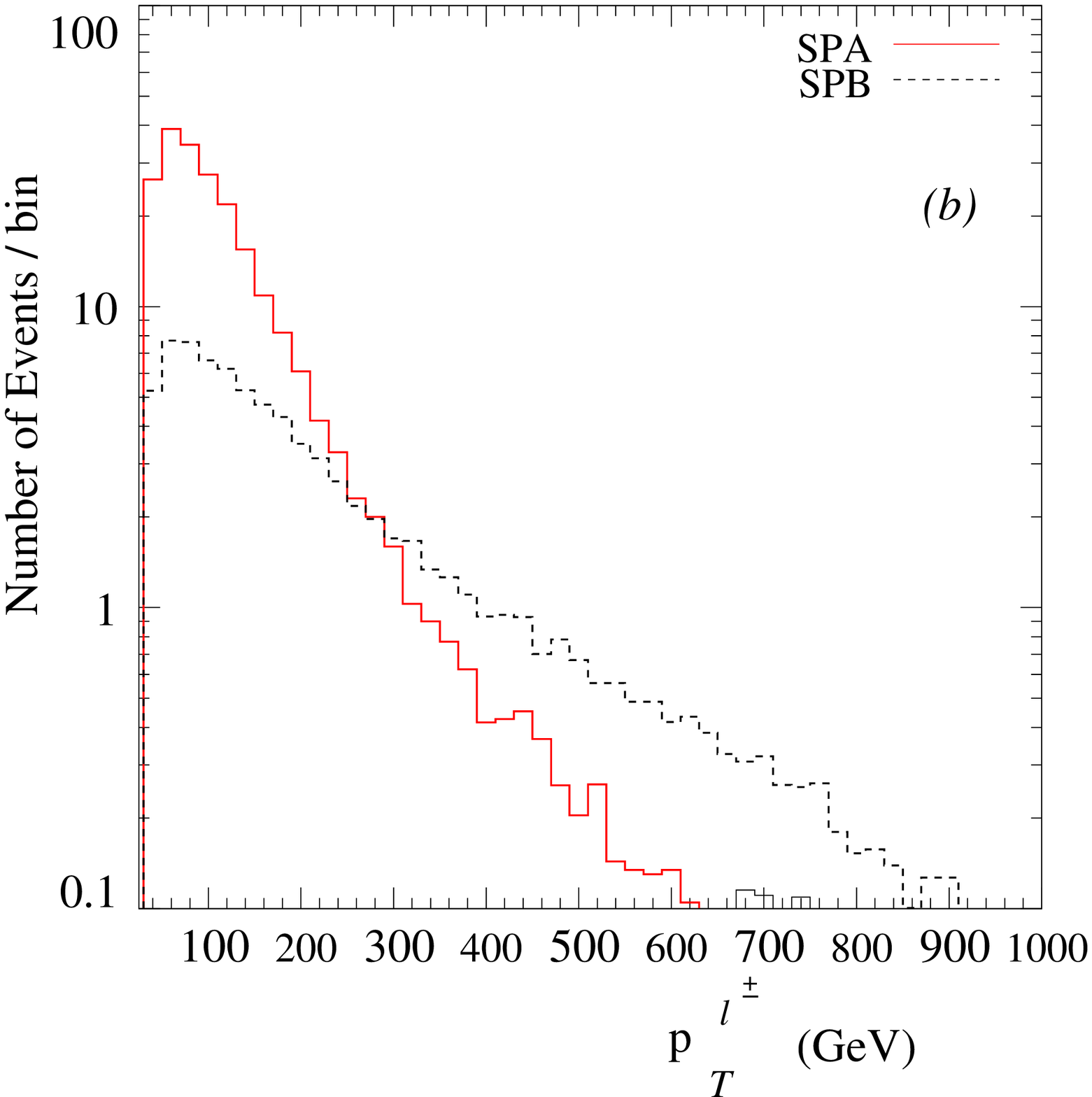}
\caption{\sl\small Binwise distribution of transverse momenta $p_T$ of the 
final state leptons with binsize of $20\ {\rm GeV}$ and integrated luminosity of
$\int{\cal L} dt = 30 fb^{-1}$. The panel (a) represents for 2-body ({\bf S2}) 
decay whereas panel (b) stands for 3-body ({\bf S3}) case.} 
\label{pT4l}
\end{center}
\end{figure}

In Fig.~\ref{pT4l}(a) and \ref{pT4l}(b) we plot the binwise distributions of 
the transverse momenta of the final state leptons for {\bf S2} and {\bf S3}, 
respectively. Since the same-sign leptons would be hard to distinguish based on
their origin (from $\Dt$ or $\tilde{\ell}_{i}$) for {\bf S2}, we prefer to plot 
the average transverse momentum of the same flavor leptons. Theoretically,
one expects leptons coming from the primary decay of $\Dt$ to be much harder
than the ones coming from intermediate slepton decay, 
$\tilde{\ell}_{i}^\pm \to \ell_i^\pm \N0_1$ for {\bf S2}. The hardness of the 
leptons, when the $\Dt$ decays through the 2-body channel, is clearly dictated 
by the mass differences between the $\Dt$, the sleptons and the LSP. 
Though this distinction is not possible at the LHC, one can understand
the larger total cross 
section for {\bf SPB (S3)} as compared to {\bf SPB (S2)}, because more soft 
leptons would be expected in the case of 2-body decays. Thus, the $p_T$ cut 
on the charged leptons has a stronger effect on the signal for {\bf SPB (S2)}. 
A quick look at  Fig.~\ref{pT4l}(a), where we plot the $p_T$ for {\bf S2} for 
both sample points, and \ref{pT4l}(b), which shows the distribution 
for {\bf S3}, indicates that one finds more events at large $p_T$ in 
Fig.~\ref{pT4l}(a) (2-body decay). This effect is due to the much harder 
leptons coming from the primary decay of the heavy  $\Dt$.
\begin{figure}[t]
\begin{center}
\includegraphics[height=2.7in,width=2.7in]{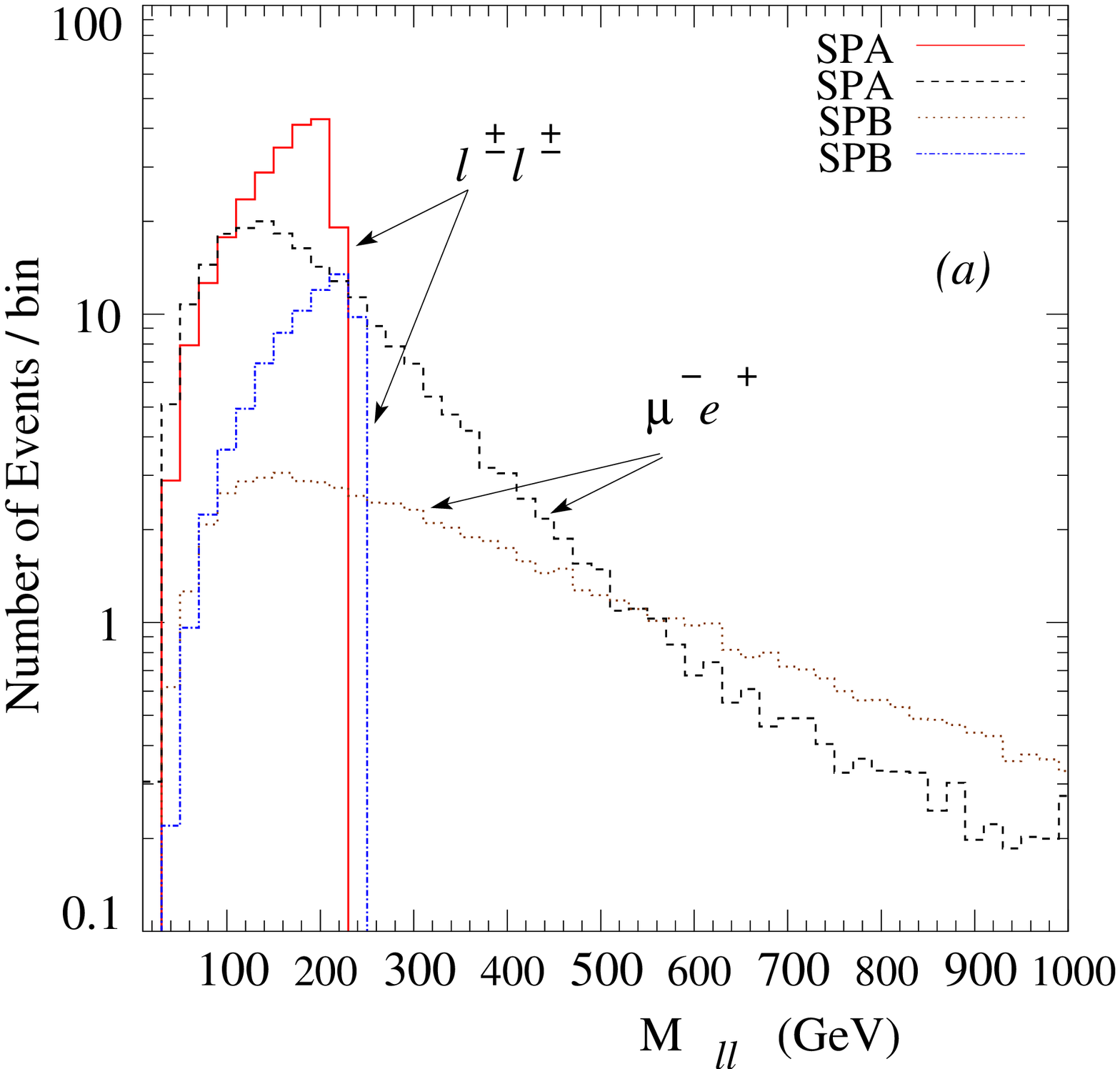} 
\includegraphics[height=2.7in,width=2.7in]{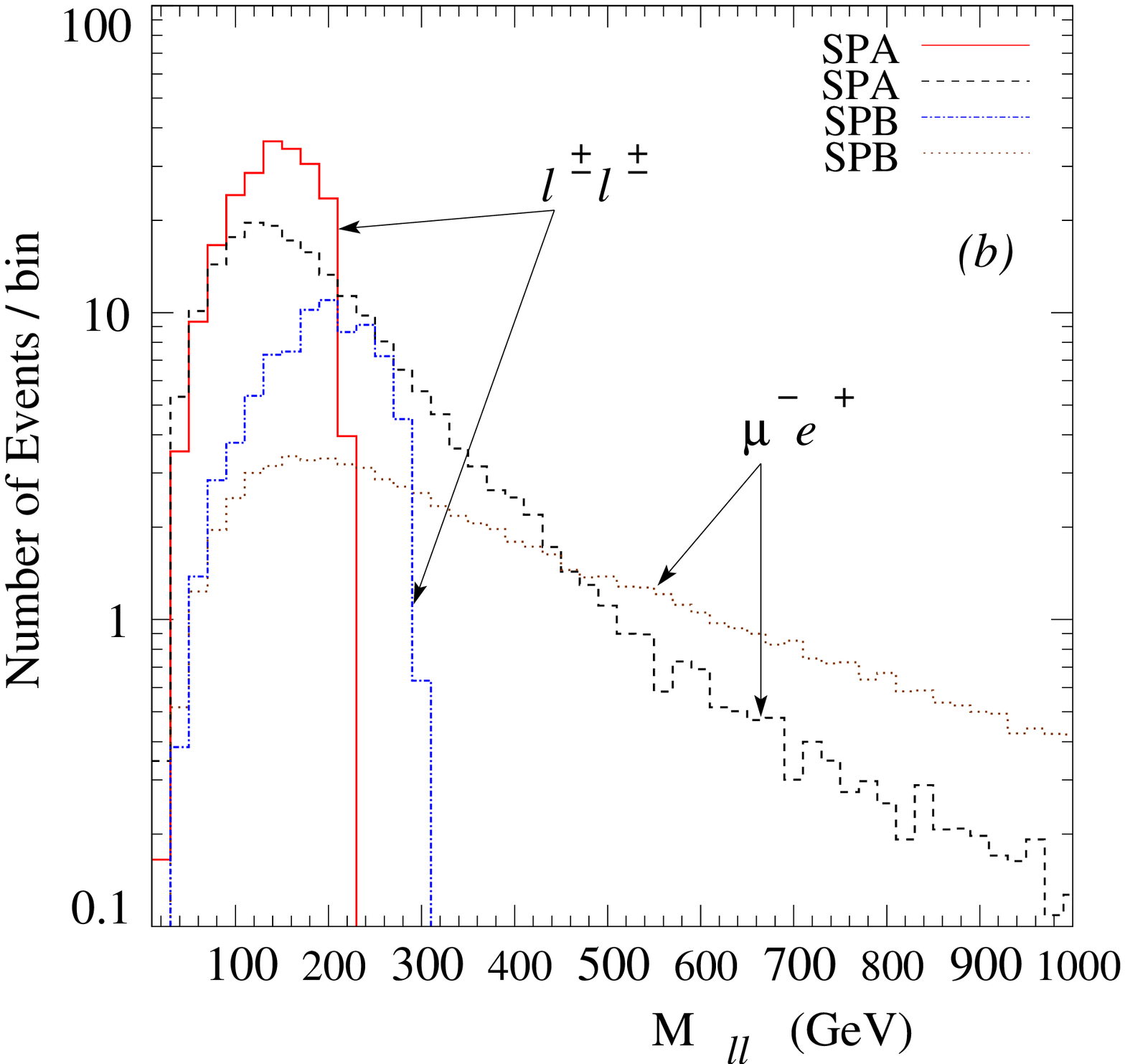}
\caption{\sl\small Binwise invariant mass distribution of lepton pairs with binsize of $20\ {\rm GeV}$ and integrated luminosity of
$\int{\cal L} dt = 30 fb^{-1}$. The panel (a) represents  the 2-body ({\bf S2}) case, and panel (b) does the 3-body ({\bf S3}) case.} 
\label{MM4l} 
\end{center}
\end{figure}

In Fig.~\ref{MM4l}(a) and \ref{MM4l}(b) we plot the binwise distributions of 
the invariant masses of the lepton pairs for {\bf S2} and {\bf S3},
respectively. These plots manifestly show differences between the SSSF and OSDF 
lepton pairs in regard to their invariant mass distributions. Indeed,
the SSSF lepton pairs exhibit a sharp kinematic edge in their $M_{\ell\ell}$
distributions whereas the OSDF lepton pairs do not. The reason,  also
mentioned when discussing Fig.~\ref{DR4l} above, is that SSSF lepton pairs 
originate from the cascade decay of the same $\Dt$. Since dilepton invariant 
mass does not change under boosts, this edge can be well-approximated for 
both {\bf S2} and {\bf S3} by the formula (in the rest frame of the decaying 
particle) 
\be
M_{\ell^\pm\ell^\pm}^{max} =
\sqrt{M_{\Dt}^2 + M_{\N0_1}^2 - 2 M_{\Dt} E_{\N0_1}}\,\,,
\label{eq:invmass}
\ee
where $E_{\N0_1}$ is the energy of the LSP. This formula yields an 
edge in the invariant mass distribution of the SSSF lepton pairs at the bin 
around $M_{\ell^\pm\ell^\pm} = M_{\Dt} - M_{\N0_1}$ for both the 
{\bf SPA} and {\bf SPB} points in the case of the 3-body decay of $\Dt$ ({\bf S3}),
as can be seen in \ref{MM4l}(b). This corresponds to the situation when the LSP
is produced at rest in the frame of $\Dt$. 
For the case {\bf S2} the situation is different, as the energy of the LSP
also depends on the mass of the slepton when the $\Dt$ decays via on-shell 
slepton ({\bf S2}). In this case the invariant mass distribution of the 
SSSF lepton pairs exhibits an edge at a different bin compared to {\bf S3}, 
as shown in Fig.~\ref{MM4l}(a) and its location is given by the formula
\be
M_{\ell^\pm\ell^\pm}^{max} = 
M_{\Dt}\sqrt{1-\left(\frac{m_{\tilde{\ell}}}{ M_{\Dt}}\right)^2}
\sqrt{1-\left(\frac{M_{\N0_1}}{m_{\tilde{\ell}}}\right)^2}
\label{eq:edgeS2}
\ee 
The edge in the SSSF dilepton invariant mass distribution yields a clear hint 
of a $\D L=2$ interaction and a doubly-charged field in the underlying model 
of `new physics'. The distributions of the OSDF dileptons exhibit
no such edge at all since in this case the two leptons originate from the 
decays of the oppositely-charged, pair-produced  $\Dt$s.

\begin{figure}[ht]
\begin{center}
\includegraphics[height=2.7in,width=2.7in]{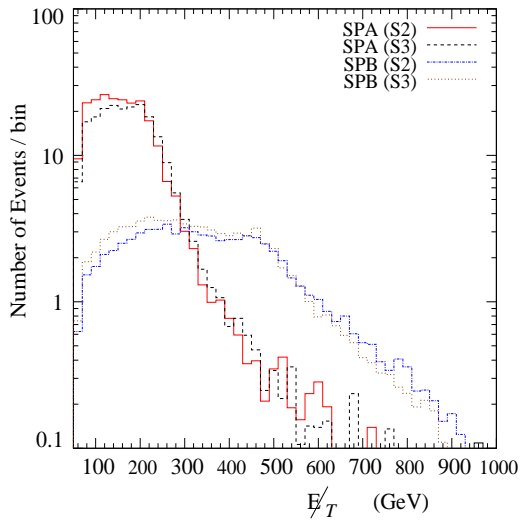} 
\caption{\sl\small Binwise distribution of the missing transverse energy
of the signal with binsize of $20\ {\rm GeV}$ and integrated luminosity of
$\int{\cal L} dt = 30 fb^{-1}$. }
\label{ETmiss4l}
\end{center}
\end{figure}

In Fig.~\ref{ETmiss4l} we plot the binwise distribution of the missing 
transverse energy for all the cases under consideration. The heavier 
neutralinos in {\bf SPB} yield more events at larger missing transverse 
energy, as expected. 


\subsection{Associated productions of doubly-charged Higgsinos and Charginos}
\begin{figure}[b]
\begin{center}
\hspace*{-0.8 in}\includegraphics[height=2.8in,width=3.4in]{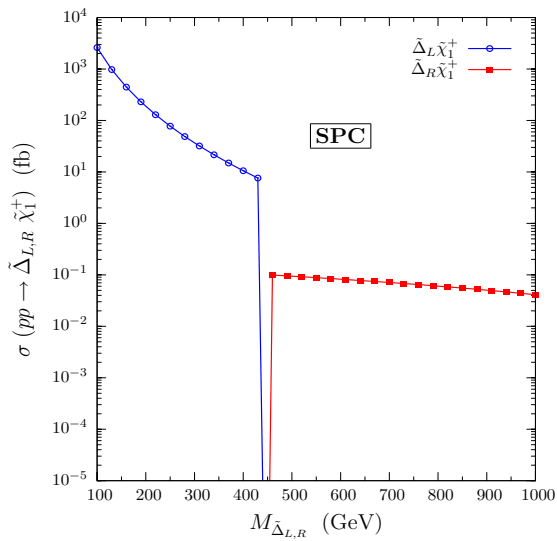} 
\caption{\sl\small The cross sections for associated productions of 
$\Dt_{L,R}$ and $\tilde{\chi}_1^{\pm}$ in the LRSUSY model at LHC. The model
parameters are as in {\bf SPC} in Table \ref{susyin}, except that  
$M_{\tilde \Delta^{--}} \equiv \mu_3$ is varied from $100\ {\rm GeV}$ up 
to $1\ {\rm TeV}$.}
\label{csecsingle}
\end{center}
\end{figure}

In this section we study productions and decays of doubly-charged Higgsinos
in association with the lightest chargino. The process under consideration,
whose Feynman diagram is depicted in Fig.~\ref{fig:singleprod}, has the form
\be
p\, p \longrightarrow \Dm\, \Cp_1 \longrightarrow
 \left(\ell_i^-\ell_i^-\right) +  \ell_j^+ + E\slash_T , \nonumber
\label{eq:singprod}
\ee
where $\ell_i$ is not necessarily identical to $\ell_j$. As mentioned above,
this scattering process proceeds with the $s$-channel $W_{L,R}$ exchange,
and yields invariably a trilepton signal, which has long been considered 
as a signal of SUSY, in general \cite{Abbott:1997je}.

The cross section for singly-produced doubly-charged Higgsino turns out to 
be small at the sample points {\bf SPA} and {\bf SPB}, and hence, we devise 
a different benchmark point, {\bf SPC}, to maximize single production of 
left-chirality doubly-charged Higgsinos. Sampling a wide region of  LRSUSY parameter space, we could not find a significant region that enhances the single production of 
right-chirality doubly-charged Higgsino. In fact, a fine-grained scan of the 
entire parameter space, with $M_{\widetilde{\Delta}}=300\gev$, yields a maximal
cross section for the right-chirality Higgsino which is still a factor of 
three smaller than that of the left-chirality Higgsino. They both become 
negligible around $M_{\widetilde{\Delta}}=300\gev$  since therein the 
composition of the lightest chargino changes abruptly. Consequently, in this 
section we use the sample point {\bf SPC} and discuss the left-chirality 
doubly-charged Higgsino production in association with the lightest 
chargino $\tilde{\chi}_1^{+}$.

Fig.~\ref{csecsingle} shows that, for all {\bf SPC} parameter space with 
varying $\mu_3$, the left-chirality doubly-charged Higgsino produced in 
association with the lightest chargino yields a large cross section for small 
Higgsino masses, and remains appreciable for doubly-charged Higgsinos as heavy 
as  $M_{\Dt} \sim 450\gev$. For the purpose of comparison, we also include 
the cross section for the right-chirality Higgsino, which starts dominating the 
cross section for the left-chirality one  as $M_{\Dt}$ becomes larger than 
$450\gev$. One notes here that, since the chargino couplings to $\Dt_{L/R}$ 
depend on the entries in the mixing matrices of charginos, the input 
parameters in Table \ref{susyin} play a crucial role in determining the 
production cross section. Since we assume $\mu_3=300\gev$ for {\bf SPC}, 
the $3\ell+E\slash_T$ signal comes from the decay of the left-chirality 
Higgsino, only. The cross section for $p\, p \to \Dm_L\, \Cp_1$ is around 
$30-40\ {\rm fb}$ for {\bf SPC}. Based on further analysis the single 
production cross section for {\bf SPC} is quite stable against large 
variations in the other parameters of the model. Of course, this does not mean 
that the same holds for the signal cross section. 
For example, even though the $\tan\beta$ dependence of production 
cross section is very weak (as long as it does not significantly change the 
$\widetilde{\Delta}_L^-$  composition of $\Cp_1$), there is a stronger 
dependence in the decay modes, as can be seen from the couplings listed 
in Appendix.
   
As in pair-production, the $\Dm$ decays again into a pair of SSSF leptons 
and an LSP following Eq.~\ref{eq:decays}, either through the 2-body decay 
mode ({\bf S2}) or the 3-body decay mode ({\bf S3}). The three possible 
chargino decay modes are depicted in Fig.~\ref{fig:singleprod}. We find that 
the chargino has almost $100\%$ branching ratio to a neutrino and slepton 
for {\bf SPC}. Then sleptons decay as  in Eq.~\ref{eq:decays}. 
This gives a $3\ell+E\slash_T$ final state where the missing transverse 
energy is due to an undetected LSP and the neutrino. For the benchmark point 
{\bf  SPC} the signal gets all the contribution from the left-chirality state. 

The single $\Dm$ production gives rise to a trilepton signal at the LHC experiments. In the 
numerical analysis, following the same notation and same kinematic cuts as
in the previous subsection, we illustrate the case where $\ell_i=\mu$ 
and $\ell_j=e$. Thus, we know that the $e^+$ always comes from the chargino 
while the same-sign muons originate from the doubly-charged Higgsino.
\begin{figure}[t]
\begin{center}
\includegraphics[height=2.7in,width=2.7in]{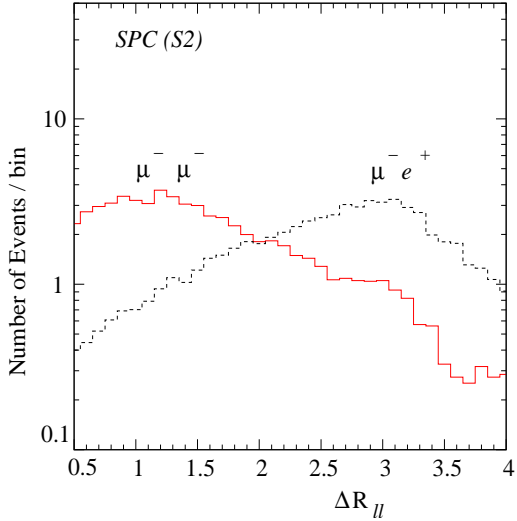} 
\caption{\sl\small Binwise distribution of $\Delta R$ with binsize 0.1 and
integrated luminosity of $\int{\cal L} dt= 30 fb^{-1}$.
}
\label{DR3l}
\end{center}
\end{figure}

The production cross section for the sample point {\bf SPC} is
\begin{itemize}
\item {\bf SPC}:
$$\sigma(\Dm_{L}\Cp_{1}) = 36.57~{\rm fb},$$
and, after imposing the kinematic cuts, the total signal cross section 
becomes
\bi
\item  {\bf S2} ~ $\sigma(2\ell^-_i \ell^+_j E\slash_T) = 2.24$ fb,
\item  {\bf S3} ~ $\sigma(2\ell^-_i \ell^+_j E\slash_T) = 2.03$ fb,
\ei
\end{itemize}
where $\ell_i = \mu$ and $\ell_j=e$. These numerical estimates hold for the 
specific choice for the final state $i.e.$ $2\mu^- + e^+ + E\slash_T$.
\begin{figure}[htb]
\begin{center}
\includegraphics[height=2.7in,width=2.7in]{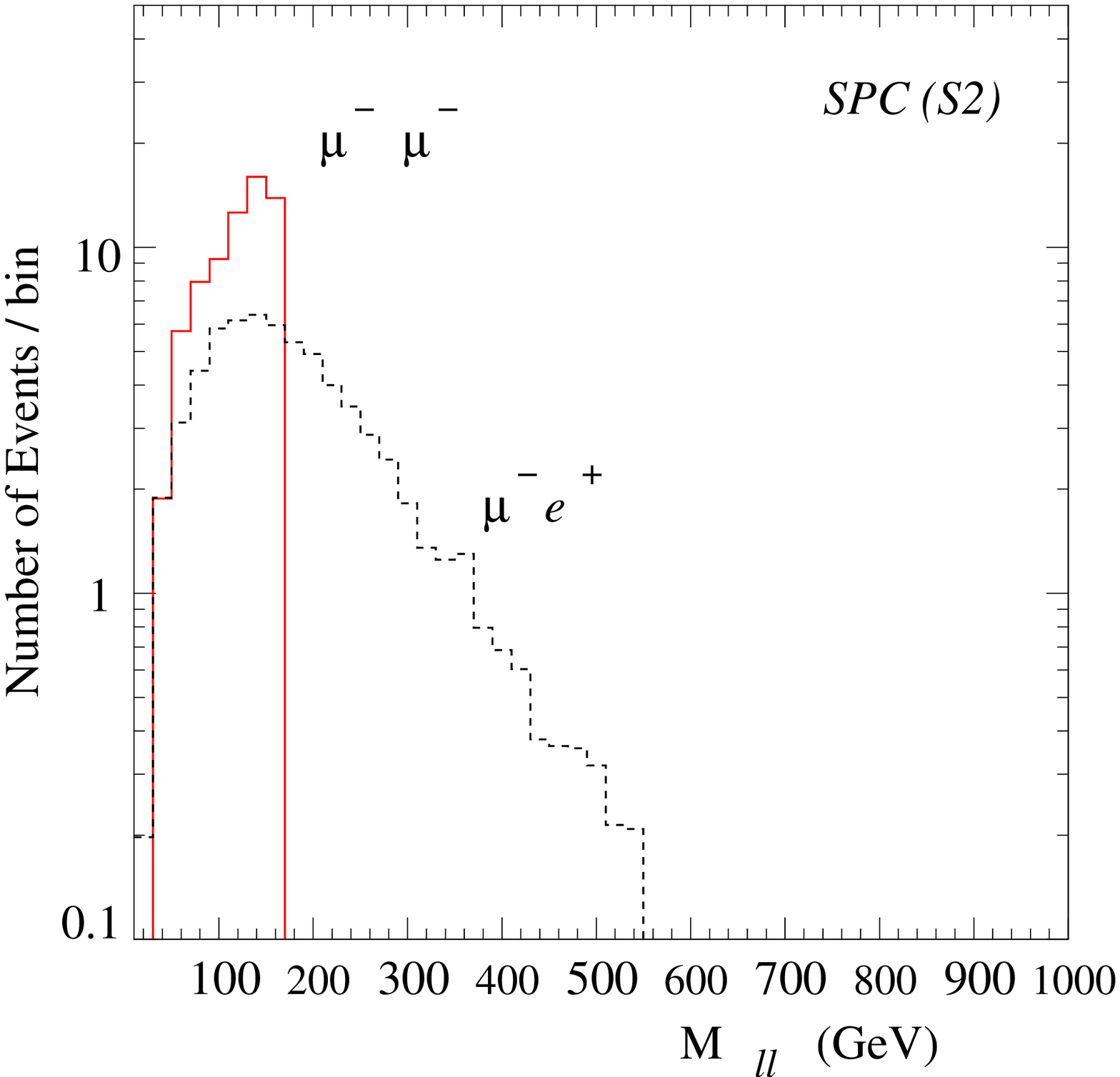} 
\includegraphics[height=2.7in,width=2.7in]{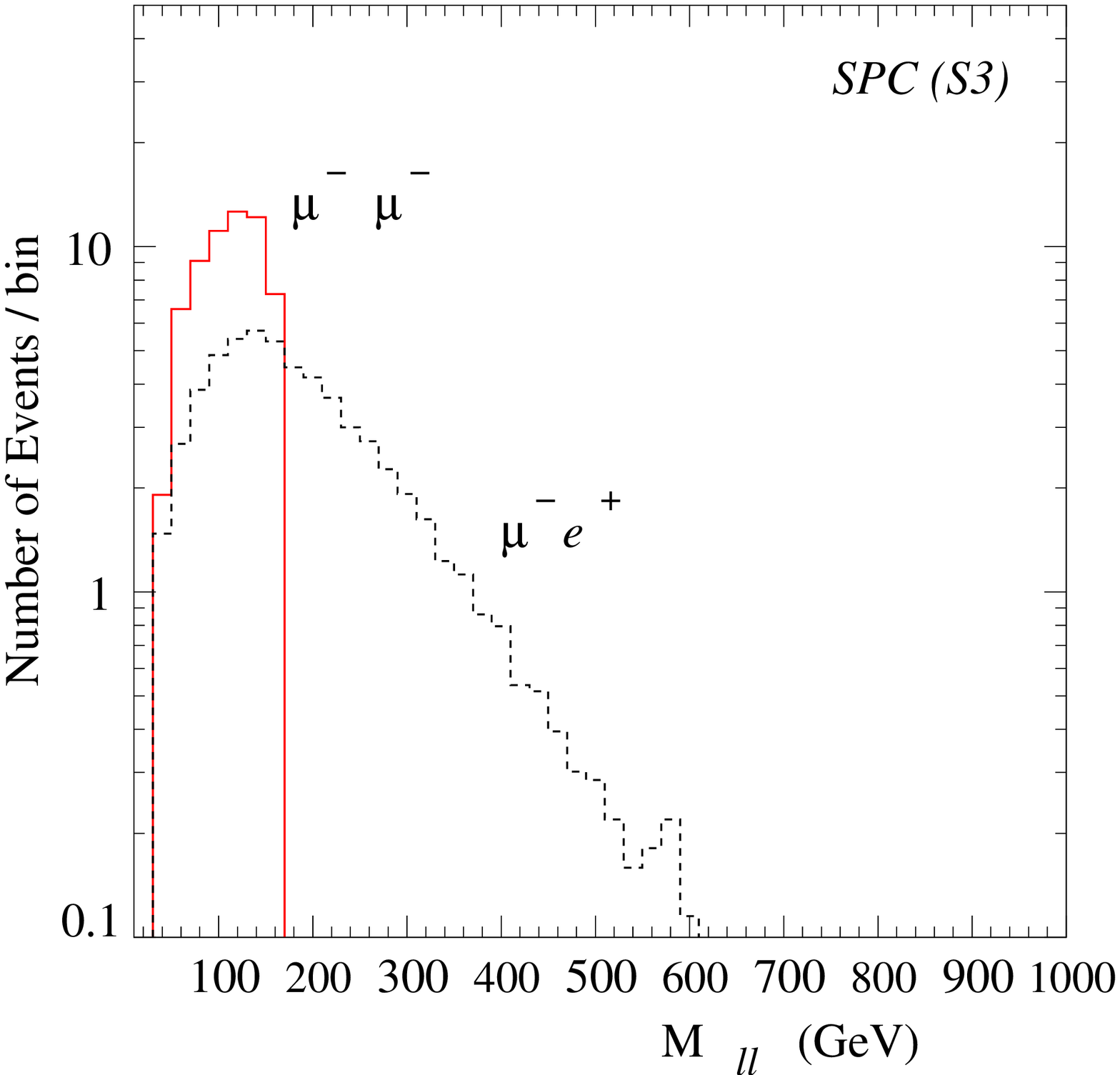}
\caption{\sl\small Binwise invariant mass distribution of pair of leptons in 
the final state with binsize $20\ {\rm GeV}$ and integrated luminosity of 
$\int{\cal L}dt=30 fb^{-1}$. }
\label{MM3l}
\end{center}
\end{figure}
\begin{figure}[h]
\begin{center}
\includegraphics[height=2.7in,width=2.7in]{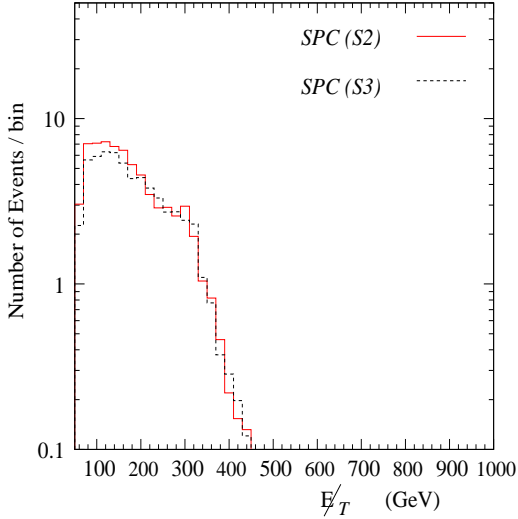} 
\caption{\sl\small Binwise distribution of missing transverse energy
for the signal with binsize $20\ {\rm GeV}$ and integrated luminosity of 
$\int{\cal L} dt = 30 fb^{-1}$. }
\label{ETmiss3l}
\end{center}
\end{figure}
\begin{figure}[htb]
\begin{center}
\includegraphics[height=2.7in,width=2.7in]{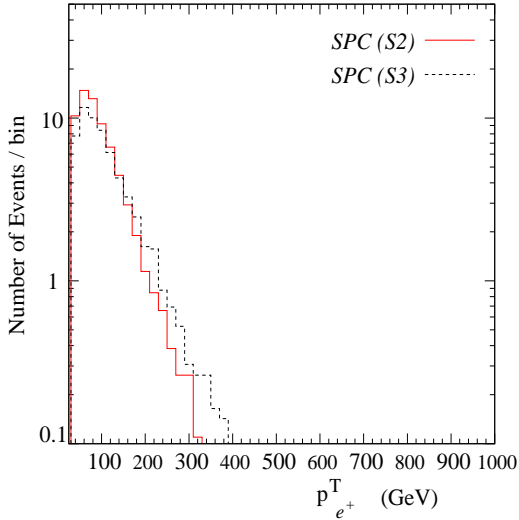} 
\caption{\sl\small Binwise distribution of transverse momentum of 
$e^+$ with binsize is $20\ {\rm GeV}$ and integrated luminosity of 
$\int{\cal L} dt = 30 fb^{-1}$. }
\label{pT_eSPC}
\end{center}
\end{figure}
\begin{figure}[htb]
\begin{center}
\includegraphics[height=2.7in,width=2.7in]{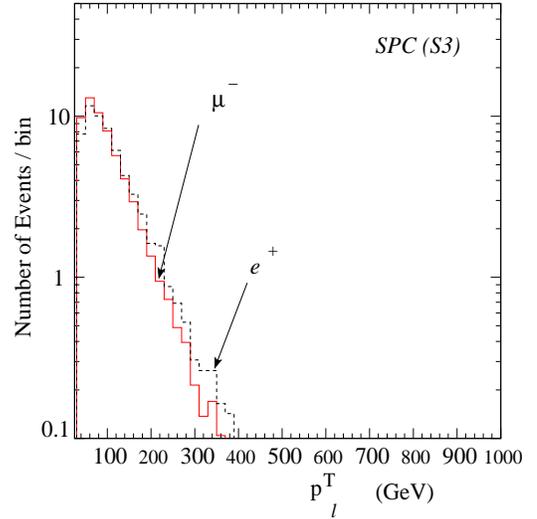} 
\caption{\sl\small Binwise distribution of transverse momentum of leptons 
for the 3-body cases with binsize is $20\ {\rm GeV}$ and integrated luminosity 
of $\int{\cal L} dt = 30 fb^{-1}$.}
\label{pT_lSPC}
\end{center}
\end{figure}

In parallel to the analysis of $4\ell+E\slash_T$ signal in previous subsection, 
we here plot various distributions in 
Figs.~\ref{DR3l},~\ref{MM3l},~\ref{ETmiss3l},~\ref{pT_eSPC} and \ref{pT_lSPC}
by considering  specifically $2\ell^-_i + \ell^+_j + E\slash_T$ signal 
with $\ell_i = \mu$ and $\ell_j = e$. Several features observed in these 
figures have already been covered by discussions in the previous subsection.
In particular, the distributions of the SSSF leptons are quite similar to 
the ones for the $4\ell+E\slash_T$ signal. This is actually expected since 
SSSF leptons are exclusively generated by decays of the doubly-charged Higgsino,
a common feature for both tetralepton and trilepton final states. 
Compared to $4\ell+E\slash_T$ signal, however, distributions for OSDF leptons 
are slightly different since the oppositely-charged electron comes exclusively 
from selectron decay, and possesses different kinematics. For example, as 
compared to $4\ell+E\slash_T$ signal, there are less events at large missing 
energy and also at large transverse momentum, for the electron from the 
chargino decay as well as the muons from the doubly-charged Higgsino decay. 
This stems from the fact that the final leptons are soft kinematically. 
It is also seen from the figures that the distributions for {\bf S2} and 
{\bf S3} are similar since the mass splitting 
$M_{\Dt}-M_{\widetilde{l}}\sim 85\gev$ is comparable to
$M_{\widetilde{l}}-M_{\N0} \sim 72\gev$.

\vskip0.2in

\section{Discussion and Conclusion}
\label{sec:conclusion}
We have studied the LHC signals of doubly-charged Higgsinos present in 
extended SUSY models such as the LRSUSY. The doubly-charged Higgs fermions in 
the spectrum are a characteristic feature of LRSUSY which can directly 
and unambiguously distinguish the model from the MSSM (and its various extensions 
like NMSSM and U(1)$^{\prime}$ models) by measuring certain leptonic events. 
We have given a detailed account of the leptonic signals originating from 
production-and-decay of $(i)$ doubly-charged Higgsino pairs and of $(ii)$ 
single doubly-charged Higgsino plus chargino. For the production mode 
$(i)$ the leptonic final state invariably involves 
$\left(\ell_i^-\ell_i^-\right) + \left(\ell_j^{+}\ell_j^{+}\right) + E\slash_T$,
that is, a pair of SSSF dileptons plus missing energy taken away by the LSP, 
$\tilde{\chi}_1^0$. On the other hand, for $(ii)$ the leptonic final state is 
composed of $\left(\ell_i^- \ell_i^-\right) + \ell_j^{+} + E\slash_T$, 
that is, a trilepton signal. Our simulation studies yield rather generically, 
for $\ell_i \neq \ell_j$, that the SSSF dileptons exhibit $a)$ a narrow 
spatial extension and $b)$ a sharp edge in dilepton invariant mass, in 
contrast to OSDF dileptons. There are additional distinctive features which 
become visible via transverse momentum/energy distributions. These 
`experimental' results provide a testing ground for an attempt to determine 
the underlying SUSY model at the ${\rm TeV}$ scale.

For a clearer view of the distinguishing power of these features, it 
proves useful to compare them with expectations of another
SUSY model such as the MSSM. Concerning the tetralepton signal 
$\left(\ell_i^-\ell_i^-\right) + \left(\ell_j^{+}\ell_j^{+}\right) + E\slash_T$
in LRSUSY, one notes that a similar signal also arises in the MSSM via 
pair-production-and-decay of the next-to-lightest neutralino $\tilde\chi_2^0$
(which is dominated by $\lambda_Z$ at least in minimal supergravity) 
with a different topology $\left(\ell_i^- \ell_i^+\right) + 
\left(\ell_j^{-} \ell_j^{+}\right) + E\slash_T$ \cite{abdullin}. Therefore, 
in contrast to leptons originating from decays of doubly-charged Higgsinos 
whose spatial distributions are shown in Fig. \ref{DR4l}, in the MSSM 
opposite-sign same-flavor (OSSF) dileptons are expected to have a 
narrow spatial extension. This and other features which follow from the plots 
in previous section enable one to distinguish between LRSUSY and MSSM in 
tetralepton signals. 

Concerning the trilepton signal $\left(\ell_i^- \ell_i^-\right) + \ell_j^{+} + E\slash_T$, one notices that a
similar signal, $\left(\ell_i^- \ell_i^+\right) + \ell_j^{+} + E\slash_T$, 
also arises in the MSSM via associated productions $\chi_2^0$ and $\tilde{\chi}_1^+$,
and their subsequent decays into leptons and $\tilde{\chi}_1^0$. As in the tetralepton
case, the two models predict different topologies for final-state leptons.
The $\tilde{\chi}_2^0$ decay gives rise to OSSF leptons and in contrast to LRSUSY
expectation depicted in Fig. \ref{DR3l}, in the MSSM OSSF leptons are expected
to have a narrow spatial extension. The trilepton signal with missing 
transverse energy has long been identified as one of the most promising 
signals of SUSY \cite{Abbott:1997je} in general. Here we see how it can be used to test for a different scenario than MSSM.

This procedure of discriminating different models of `new physics' with
lepton spectrum naturally extends to other models, not necessarily of supersymmetric 
nature. For example, in universal extra dimensions (UED), pair-production of two excited $Z$ bosons
-- the first Kaluza-Klein (KK) level $Z_1$ -- invariably leads to tetralepton
signals through the cascade decay $Z_1 \rightarrow \ell_1^- \ell^+ \rightarrow 
LKP\, \ell^- \ell^+$ of each $Z_1$ (with LKP being the lightest KK particle
whose stability is guaranteed by the KK symmetry) \cite{Cheng:2002ab}. By the same
token,  the trilepton signal follows from the associated production of charged and
neutral gauge bosons, $W_1^{\pm}\, Z_1$, and their subsequent decays into leptons and LKP. 
In terms of the event topologies, trilepton and tetralepton signals of UED are similar to 
those of the MSSM, and thus, distinguishing UED from LRSUSY is accomplished with
the same strategy used for the MSSM.

Also interesting are models with  low-scale $U(1)_{B-L}$ invariance, which
accommodate a light right-handed Majorana neutrino $N$ \cite{Khalil:2006yi}.
The pair-produced right-handed neutrinos can give rise to tetralepton signal
via $N\rightarrow \ell_i^+ W^-\rightarrow \ell_i^+ \ell_j^- \bar\nu_j$ decay. The
trilepton signal can come from associated $\ell_i\, N$ production and is
strongly suppressed. The LHC signatures of this model are similar to those
of the MSSM and UED, and SSSF lepton distributions enable one to distinguish
it from LRSUSY \cite{Huitu:2008gf}.

These case studies can be extended to a multitude of `new physics' models
at both qualitative and quantitative level. In each case, LRSUSY, whose 
spectrum consists of doubly-charged Higgsinos, is found to differ from the
rest by having SSSF proximate dileptons at the final state. Our results show  
convincingly clear that doubly-charged Higgsinos give rise to rather special 
leptonic events at the LHC, making them firmly distinguishable 
from other SUSY particles and also from particles in several other models
of physics at the ${\rm TeV}$ scale.

\acknowledgments
The work of M.F. and I.T. is supported in part by NSERC of Canada under the
Grant No. SAP01105354. The work of D.D. was supported by  
Alexander von Humboldt-Stiftung Friedrich Wilhelm Bessel-Forschungspreise and 
by the Turkish Academy of Sciences via GEBIP grant. 
K.H. and S.K.R. gratefully acknowledge the support from the Academy of
Finland (Project No. 115032). We would like to thank M.~T.~Ataol and 
P.~M.~K.~Ravuri for useful technical discussions about {\tt CalcHEP} package, 
A. Belyaev for discussions on {\tt CalcHEP}-Pythia interface,
R. Kinnunen and S. Raychaudhuri for discussions, and Goran Senjanovi{\' c} for 
enlightening e-mail exchange.

\begin{appendix}
\begin{center}
{\bf APPENDIX}
\end{center}
In this Appendix we list down all the Feynman rules necessary for analyzing
productions and decays of doubly-charged Higgsinos in the LRSUSY model.

\noindent
{\bf\underline{Scalar-Scalar-Z Boson, $\gamma$:}}
\\

$\bullet A^{\mu}~ \tilde q ~\tilde q^{\star}:\,  ~~ -ieQ_q(p_q+p_{q^*})^{\mu}$

 $
\bullet Z_L^{\mu}\, \tilde q\, \tilde q^{\star}:~~~-i\frac{g_L}{ \cos \theta_W}(T_{3q}^L-Q_f \sin^2 \theta_W) (p_q+p_{q^*})^{\mu}$

$
\bullet Z_R^{\mu}\,\tilde q\,\tilde q^{\star}:~~~~ -i\,\frac{g_R \sqrt{\cos 2\theta_W}}{ \cos \theta_W}(T_{3q}^R-\frac16 \frac{\sin^2 \theta_W}{\cos 2 \theta_W}) (p_q+p_{q^*})^{\mu}$


\noindent
\\
{\bf\underline{Scalar-Scalar-W bosons:}}
\\

$\displaystyle
\bullet W^{\mu}_L~ \tilde l_L~ \tilde \nu_L: ~ ~~~-i\frac{g_L}{\sqrt {2}}(p_l+p_\nu)^{\mu}$

$\displaystyle
\bullet W^{\mu}_R~ \tilde l_R ~\tilde \nu_R: ~ ~~~-i\frac{g_R}{\sqrt {2}}(p_l+p_v)^{\mu}$

\noindent
\\
{\bf\underline{Fermion-Fermion-W bosons:}}
\\

$\displaystyle
\bullet W^{\mu}_L ~ l~ \bar \nu:  ~~~~~~~~~~~~-i\frac{g_L}{\sqrt {2}}\gamma^{\mu}P_L$

$\displaystyle
\bullet W^{\mu}_R  ~l  ~\bar \nu:  ~~~~~~~~~~~~-i\frac{g_R}{\sqrt {2}}\gamma^{\mu}P_R$

$\displaystyle
\bullet W^{\mu}_L ~q ~\bar q':  ~~~~~~~~~~~-i\frac{g_L}{\sqrt {2}}\gamma^{\mu}P_L$

$\displaystyle
\bullet W^{\mu}_R~ q~ \bar q':  ~~~~~~~~~~~-i\frac{g_R}{\sqrt {2}}\gamma^{\mu}P_R$

$\displaystyle
\bullet W^{\mu}_L ~\tilde{\chi}_k^+ ~\tilde \Delta_L^{--}: ~~~~~~~~ ig_L\gamma^{\mu}(V^{\star}_{k5}P_L+U_{k5}P_R)$

$\displaystyle
\bullet W^{\mu}_R \tilde{ \chi}_k^+ \tilde \Delta_R^{--}: ~~~~~~~~~~~~ ig_R\gamma^{\mu}(V^{\star}_{k6}P_L+U_{k6}P_R)$

$\displaystyle
\bullet W^{\mu}_L ~ \tilde{\chi}_k^+ ~\tilde{\chi}^0_j:~~~~~~~~~ -ig_L\gamma^{\mu}(L^{L}_{jk}P_L+L^{R}_{jk}P_R)$

$\displaystyle
\bullet W^{\mu}_R ~ \tilde{\chi}_k^+ ~\tilde{\chi}^0_j: ~~~~~~~~~~ -ig_R\gamma^{\mu}(R^{L}_{jk}P_L+R^{R}_{jk}P_R)$
with the matrix elements given in terms of chargino and neutralino mixing matrices as
\begin{eqnarray}
L^L_{jk}\!\!&=&\!\!-N^{\star}_{k1}V_{j1}+\frac{1}{\sqrt{2}}N^{\star}_{k5}V_{j4}+ +N_{k6}^{\star}V_{j5}+\frac{1}{\sqrt{2}}N^{\star}_{k11}V_{j3}
\nonumber \\
L^R_{jk}\!\!&=&\!\!-U^{\star}_{j1}N_{k1}-\frac{1}{\sqrt{2}}U^{\star}_{j4}N_{k4}+N_{k7}^{\star}V_{j5}-\frac{1}{\sqrt{2}}U_{j4}^{\star}N_{k10}
\nonumber \\
R^L_{jk}\!\!&=&\!\!-N^{\star}_{k2}V_{j2}+\frac{1}{\sqrt{2}}N^{\star}_{k5}V_{j4}+N^{\star}_{k8}V_{j6}+\frac{1}{\sqrt{2}}N^{\star}_{k11}V_{j3}
\nonumber \\
R^R_{jk}\!\!&=&\!\!-U^{\star}_{j2}N_{k2}-\frac{1}{\sqrt{2}}U^{\star}_{j3}N_{k4}+U^{\star}_{j6}N_{k9}-\frac{1}{\sqrt{2}}U_{j4}^{\star}N_{k10}\nonumber
\end{eqnarray}

\noindent
{\bf\underline{Fermion-Fermion-Z Boson, $\gamma$:}}

$\displaystyle
\bullet \gamma^{\mu} \tilde\Delta_{L,R}^{--} \bar{\tilde\Delta}_{L,R}^{--}:~~~~~~~~~~~2i e \gamma^{\mu}$

$\displaystyle
\bullet Z_L^{\mu} \tilde\Delta_{L}^{--} \bar{\tilde\Delta}_{L}^{--}:~~~~~~~~~~~i\frac{g_L \cos 2 \theta_W}{\cos \theta_W} \gamma^{\mu}$

$\displaystyle
\bullet Z_L^{\mu} \tilde\Delta_{R}^{--} \bar{\tilde\Delta}_{R}^{--}:~~~~~~~~-i\frac{2g_L \sin^2 \theta_W}{\cos \theta_W} \gamma^{\mu}$

$\displaystyle
\bullet Z_R^{\mu} \tilde\Delta_{L}^{--} \bar{\tilde\Delta}_{L}^{--}:~~~~~~~~~~~i\frac{g_L \sqrt{\cos 2 \theta_W}}{\cos \theta_W} \gamma^{\mu}$

$\displaystyle
\bullet Z_R^{\mu} \tilde\Delta_{R}^{--} \bar{\tilde\Delta}_{R}^{--}:~~~~~~~~-i\frac{g_L (1-3 \sin^2\theta_W)}{\cos \theta_W \sqrt{\cos 2 \theta_W}} \gamma^{\mu}$

\clearpage
\begin{widetext}
\noindent
{\bf\underline{Fermion-Fermion-Scalar Fermion:}}
\\

$\bullet \tilde \Delta_L^{--} ~\tilde l~ l: ~~ ~~~~-2h_{ll}\mathcal{C}^{-1}P_L$

$\bullet\tilde \Delta_R^{--}~ \tilde l~ l:  ~~~~~~-2h_{ll}\mathcal{C}^{-1}P_R$

$\bullet~\tilde \Delta_L^{-} ~\tilde l~ \nu:  ~~~~~~~~~h_{ll}\mathcal{C}^{-1}P_L$

$\bullet~\tilde \Delta_L^{-}~ l \tilde \nu:  ~~~~~~~~~~ h_{ll}\mathcal{C}^{-1}P_L$

$\bullet~\tilde \Delta_R^{-}~ \tilde l ~\nu:  ~~~~~~~~~h_{ll}\mathcal{C}^{-1}P_R$

$\bullet~\tilde \Delta_R^{-}~ l ~\tilde \nu:  ~~~~~~~~~h_{ll}\mathcal{C}^{-1}P_R$

$\displaystyle
\bullet~\tilde{\chi}_k^0~ \tilde l~ \bar l:~~~~~-i\left\{ \left[\sqrt{2} g_L \left ( \frac12N_{k1}- \frac12 ( \frac{\cos 2 \theta_W}{\cos^2 \theta_W}+\tan^2\theta_W) N_{k2} -\frac{\sin \theta_W \sqrt{\cos 2\theta_W}}{\cos^2 \theta_W}N_{k3}  +\frac{m_l}{2 M_W \cos \beta}N_{k5}\right )\right] P_L\right. 
\\
\left.~~~~~~~~~~~~~~~~~~ -\left[\sqrt{2}g_R \left ( (\frac{\cos 2 \theta_W}{2\cos^2 \theta_W}-\tan^2 \theta_W)N_{k2}- \frac{\sin \theta_W \sqrt{\cos 2\theta_W}}{\cos^2 \theta_W}N_{k3}
+\frac{m_l}{2 M_W \cos \beta}N_{k5}^{\star}\right ) \right]P_R\right \}$

$\displaystyle
\bullet~\tilde{\chi}_k^0~ \tilde \nu~ \bar \nu:~~~~-i\left\{ \left[\sqrt{2} g_L \left ( \frac12N_{k1}+ \frac12 ( \frac{\cos 2 \theta_W}{\cos^2 \theta_W}-\tan^2\theta_W) N_{k2} -\frac{\sin \theta_W \sqrt{\cos 2\theta_W}}{\cos^2 \theta_W}N_{k3}  +\frac{m_l}{2 M_W \cos \beta}N_{k5}\right )\right] P_L\right. 
\\
\left.~~~~~~~~~~~~~~~~~~ -\left[\sqrt{2}g_R \left (\frac{\cos 2 \theta_W}{2\cos^2 \theta_W}N_{k2}- \frac{\sin \theta_W \sqrt{\cos 2\theta_W}}{2\cos^2 \theta_W}N_{k3}
+\frac{m_l}{2 M_W \cos \beta}N_{k5}^{\star}\right ) \right]P_R\right \}$
\end{widetext}
\end{appendix}

\end{document}